\documentclass[11pt, a4paper]{article}
\usepackage[utf8]{inputenc}
\usepackage{amsmath}
\usepackage{amssymb}
\usepackage{multirow}
\usepackage{booktabs}
\usepackage{natbib}
\usepackage{hyperref}
\usepackage{amsthm}
\usepackage{verbatim}
\usepackage{graphicx}
\usepackage{subcaption}
\usepackage{enumitem}
\usepackage{textcomp}
\usepackage{mathtools}
\usepackage{cancel}
\usepackage{placeins}
\usepackage{algorithmic,algorithm}

\usepackage{PRIMEarxiv}

\usepackage[utf8]{inputenc} 
\usepackage[T1]{fontenc}    
\usepackage{hyperref}       
\usepackage{url}            
\usepackage{booktabs}       
\usepackage{amsfonts}       
\usepackage{nicefrac}       
\usepackage{microtype}      
\usepackage{lipsum}
\usepackage{graphicx}
\usepackage{blindtext}
\usepackage{amsthm,amsmath,amsfonts,amssymb}
\usepackage{natbib}

\usepackage{subcaption}
\usepackage{comment}
\usepackage{tikz}
\usepackage{float}
\usepackage{multirow}
\usepackage{float}
\usepackage{dsfont}
\usepackage{booktabs}
\usepackage{bigstrut}
\usepackage[export]{adjustbox}
\usepackage{siunitx} 

\newcommand\independent{\protect\mathpalette{\protect\independenT}{\perp}}
\def\independenT#1#2{\mathrel{\rlap{$#1#2$}\mkern2mu{#1#2}}}
\theoremstyle{plain}
\theoremstyle{remark}
\DeclareMathOperator*{\argmin}{arg\,min}

\title{Probabilistic Rainfall Downscaling: \\Joint Generalized Neural Models with Censored Spatial Gaussian Copula}

\author{
  David Huk, Rilwan A. Adewoyin\thanks{Also affiliated with Department of Computer Science and Engineering, Southern University of Science and
Technology, Shenzhen, China.},  Ritabrata Dutta\\
  Department of Statistics\\
        University of Warwick\\
  \texttt{\{david.huk, rilwan.adewoyin, ritabrata.dutta\}@warwick.ac.uk}}

\begin{document}
\maketitle

\begin{abstract}

A novel approach for generating conditional probabilistic rainfall downscaling at finer scales from deterministic weather variables at coarser scales with temporal and spatial dependence is introduced. A two-step procedure is employed. Firstly, marginal location-specific distributions are jointly modelled conditional on the deterministic coarse weather variables. Secondly, a spatial dependency structure is learned to ensure spatial coherence among these distributions.
To learn marginal distributions over rainfall values, we introduce joint generalised neural models that expand generalised linear models with a deep neural network architecture to jointly fit parameters of the distributions.
The spatial dependency structure is modelled using a censored latent Gaussian copula leveraging the underlying spatial structure. We construct a distance matrix between locations, transformed into a correlation matrix by a Gaussian Process Kernel depending on a small set of parameters. To estimate these parameters, we propose a general framework for the estimation of latent Gaussian copulas employing scoring rules as a measure of divergence between distributions. 
Uniting our two contributions, namely the joint generalised neural model and the censored latent Gaussian copulas into a single model, our probabilistic approach provides downscaled rainfall. We demonstrate its efficacy using a large UK data set, outperforming existing methods.
\end{abstract}

\section{Introduction}
In an era of escalating climate change, the global community is witnessing an intensifying frequency of high precipitation events. As per the United Nations' report, flooding emerged as the most prevalent weather-induced disaster over the two decades culminating in 2015, impacting 2.3 billion individuals and resulting in an economic toll of approximately 1.89 trillion \citep{2015UNreportthehumancostof}. In response to the mounting fiscal and societal risks associated with heavy precipitation and flooding, policy makers are increasingly leveraging large-scale numerical weather predictors (NWPs) \citep{randall2007climate,jung2010ecmwf} in conjunction with flood models (FMs) \citep{qi2021review}.\par
One of the main shortcomings of NWPs is their coarse resolution. State-of-the-art NWPs act on horizontal scales of 50 km to 100 km, which is insufficient to capture local precipitation. But to accurately predict possible future flooding scenarios, FMs do need an accurate prediction of precipitation on a fine grid (for example, a 10 km grid) for all locations simultaneously in a large area (such as all of the UK). To provide local-scale precipitation predictions, it is common practice to downscale coarse NWP forecasts to a finer grid with the use of post-processing methods. 
Similar to the works by \cite{Rover, Vandal2018, 2021Trunet,xiang2022novel} and \cite{wang2023customized}, one of our main aims is to perform statistical \emph{downscaling} of rainfall, that is predicting high-resolution precipitation from low-resolution weather variables. 

Furthermore, one would like to have uncertainty quantification about possible rainfall levels over a large area in order to assess various precipitation-related risks (e.g. risks of different flooding scenarios) \citep{schoof2013statistical}, which is not possible with deterministic downscaling, motivating our model to provide \emph{probabilistic} downscaling of precipitation. Predicting a distribution for rainfall rather than merely predicting the expected value can also rectify the deficiency in the simulation of extreme precipitation \citep{watson2022machine}, as point estimates are inherently biased towards the more commonly observed low rainfall events. To provide accurate estimates of risks of different flooding scenarios through flood models, we need to be able to predict and sample from the joint distribution of precipitation from a large area in a \emph{spatially-coherent} manner rather than only providing probabilistic prediction of precipitation in a location-specific manner. Unlike implicit generative Machine Learning based approaches such as \cite{ravuri2021skilful,Harris2022,chen2022generative,duncan2022generative} and \cite{price2022increasing} which do not have access to the distribution or cannot output precisely zero-valued observations, we would like to obtain an explicit model for rainfall quantities capable of predicting exact zero amounts, thus retaining the interpretability of the model's outputs.

Hence our \textbf{main aim} is the development of a downscaling method for the reliable and explainable mapping of large-scale atmospheric variables from coarse weather or climate models to \emph{spatially-coherent high-resolution probabilistic predictions of precipitation}.

To this aim, we introduce a pioneering two-component probabilistic model for the downscaling of rainfall. The first component comprises a novel joint generalised neural model designed to learn location-specific explicit marginal distributions conditioned on spatio-temporal covariates through a novel architecture. (Note that here, ``joint” refers to the estimation of marginal density parameters being done jointly, not
 to be confused with the estimation of a joint density). The second component is a censored latent Gaussian copula approach, specifically designed for censored data, which introduces spatial dependency into the location-specific dwonscaling of precipitation. When combined, these two components generate spatially and temporally coherent joint distributions of rainfall. We provide a diagram of our model, termed Cens-JGNM, in Figure \ref{fig:diagram}. We apply our model to an extensive UK rainfall data set and juxtapose its performance with established benchmark methods. Our approach has important applications, including:
\begin{itemize}
    \item The extension of univariate forecasts to joint distributions with coherent spatial dependence. Many models in the Geoscience community exist for reliable location-wise prediction of censored observations. Our methodology offers an easy extension of the existing models to simultaneously predict variables of interest at multiple stations while taking into account both the censoring of data and the inherent spatial dependency. 
    
    \item The improvement of rainfall predictions (for single or multiple locations) at a local level by enhancing their resolution with our approach, taking additional high-resolution information into account. Rainfall predictions are notably more challenging and have a shorter prediction horizon compared to other atmospheric variables due to the highly variable and localized nature of precipitation events. Factors such as complex cloud microphysics, rapid changes in atmospheric moisture, and small-scale convective processes contribute to this decreased predictability as they evolve at a scale significantly finer than our input data. Consequently, our approach aims to extend the temporal reach of rainfall forecasts beyond that of standard weather or climate models by exploiting correlations between more predictable atmospheric features and precipitation. Furthermore, our explicit modelling of distributions ensures explainability and captures the uncertainty inherent in precipitation downscaling.

    \item The computationally efficient generation of downscaled total precipitation outputs that enhance the accuracy of total precipitation predictions from leading NWP models, such as ERA5, which are known to exhibit significant bias \citep{https://doi.org/10.1002/qj.4351}. This problem also extends to data driven approaches which rely on NWP output as training data and are prone to learning the biases present. As such, prominent data driven approaches have excluded total precipitation from outputs or diagnostics in  \citep{lang2024aifs,lam2023learning}.

\end{itemize}

The remainder of this paper is structured as follows: Section \ref{sec:data_stat} introduces the UK meteorological and rainfall data set employed in this study (ERA5 and E-OBS for training and IFS for benchmarking) along with our two-component approach. Section \ref{sec:marginal_deep} details the application of joint generalised neural models as a method for learning marginal distributions. Section \ref{sec:cens_copula} explains how censored latent Gaussian copulas can extend marginal densities into a joint prediction with the correct spatial dependence in the presence of censored observations. Section \ref{sec:real_data} showcases the results of censored Gaussian copulas fitted with the joint generalised neural model marginal distributions to the UK data set. Connections with existing work are established in Section \ref{sec:lit_rev} with this study's conclusion presented in Section \ref{sec:ccl}.

\section{Data and Statistical Problem}
\label{sec:data_stat}

We begin by providing a description of the data in Section \ref{sec:data}, followed by an explanation of our statistical problem and approach in Section \ref{sec:stat}. Notation for the main text is included in Table \ref{tab:notation} with deviations from this indicated in the relevant Sections.

\begin{table}[ht]
    \centering
    \renewcommand{\arraystretch}{1.3} 
    \begin{tabular}{|p{1.3cm} p{5.4cm}|!{\vrule width 1.3pt} p{1.3cm} p{5.4cm}|}
        \hline
        \textbf{Symbol} & \textbf{Description} & \textbf{Symbol} & \textbf{Description} \\ 
        \hline
        $\mathcal{I}$ & Set of location $\{1,\ldots,n\}$& $i$ & Individual location \\ 
        \hline
        $\mathcal{T}$ & Set of days & $t$ & Individual day \\ 
        \hline
        $\mathcal{Y}_s$ & Target rainfall over all locations in $\mathcal{I}$ for day $s$ & $y_{i,s}$ & Target rainfall at location $i$ and day $s$\\
        \hline
        $\mathcal{Y}_s^*$ &
        Latent variable vector over all locations $\mathcal{I}$ for day $s$ &
        $y_{i,s}^*$ &
        Latent variable at location $i$ and day $s$\\
        \hline
        $\mathcal{X}_{i,s}$ & Weather variables relevant for location $i$ and day $s$& $\boldsymbol{x}_{i,s}$& Subset of predictors $ \subseteq\boldsymbol{\mathcal{H}}_s$ relevant to predict $y_{i,s}$\\ 
        \hline
        $\mathcal{H}_s$ & History of $\mathcal{X}_{i,s}$ over all $i$, and for days relevant to day $s$ & $\mathcal{D}$& Distance matrix $\in\mathbb{R}^{n\times n}$ of spatial information for all locations.\\ 
        \hline
        $F_{i,s}$ &
        (Conditional) distribution of rain at location $i$ for day $s$ &
        $F_{i,s}^*$ &
        Latent (conditional) distribution at location $i$ for day $s$ \\
        \hline
        $f_{i,s}$ &
        (Conditional) density of rain at location $i$ for day $s$&
        $f_{i,s}^*$ &
        Latent  (conditional) density at location $i$ for day $s$\\
        \hline
        $\pi$ &
        Joint (conditional) rainfall density over $\mathcal{I}$ for day $s$ &
        $\pi^*$ &
        Latent joint (conditional) density over $\mathcal{I}$ for day $s$.\\
        \hline
        $\mathbf{c}$ &
        Gaussian copula density in $n$ dimensions&
        $\mathbf{c}^*$ &
        Latent copula density of quantiles from $F_{i,s}^*$ over $\mathcal{I}$\\
        \hline
        $\Phi$ &
        Standard Gaussian distribution function&
        $\phi$ &
        Standard Gaussian density function\\
        \hline
        $z_{i,s}$ &
        Observed data on Gaussian scale $\Phi^{-1}\left(F_{i,s}(y_{i,s}|\boldsymbol{\mathcal{H}}_s)\right)$&
        $z_{i,s}^{*}$ &
        Latent data on Gaussian scale $\Phi^{-1}\left(F_{i,s}^{*}(y_{i,s}^{*}|\boldsymbol{\mathcal{H}}_s)\right)$\\
        \hline
    \end{tabular}
    \caption{Notations used throughout the main text.}
    \label{tab:notation}
\end{table}

\subsection{Meteorological Data}
\label{sec:data}
\paragraph{ERA5 and E-OBS data:}Our model aims to downscale weather variables from the ERA5 reanalysis data set \citep{ERA5} on a coarse grid to a joint distribution of rainfall across multiple locations on a finer grid as given by the E-OBS rainfall data \citep{Eobscite}. Table~\ref{tab:datasources} summarizes the data sources involved in this work. \par

The ECMWF employs a physics based weather simulator named the Integrated Forecasting System (IFS) for operational weather predictions and to generate the ERA5 \citep{ERA5} reanalysis data, which serves as input data for our work. 
We use a selection of model fields, which can be resolved reasonably well by
typical climate models, from the ERA5 reanalysis dataset (Hersbach et al., 2020) as input variables for our model, as we hope to extend our model to climate model inputs in future works. Formally, at 6‐hourly intervals $j \in \{1,2,3,4\}$ on day $t \in \mathcal{T} \subset \mathbb{N}$ and for locations $i \in \{1, \ldots, n\} = \mathcal{I}$, we are given predictors 
\[
\boldsymbol{\mathcal{X}}_{i,jt} = (\mathcal{X}_{i,jt}^{(1)}, \ldots, \mathcal{X}_{i,jt}^{(6)}),
\]
representing six weather variables from the ERA5 reanalysis: specific humidity, air temperature, geopotential height (500 hPa), the eastward and northward components of wind velocity (850 hPa), and total column water vapour. These variables are available four times per day at a resolution of approximately 5/9° in longitude and 5/6° in latitude (corresponding to roughly 50 km grid spacing over Wales) as produced by the IFS-based reanalysis. 

For each day, daily means of these smoothed model fields are computed. However, because our outcome variables—the 24‐hour precipitation accumulations (in millimetres) from the E-OBS dataset—are recorded on a 0.1° longitude-latitude grid, it is necessary to interpolate the input fields to the same high-resolution grid. In practice, the original coarse grid (which covers a spatial extent roughly corresponding to a 20×21 grid) is regridded using a bivariate spline interpolation scheme on a rectangular mesh to produce a finer 100×140 grid with 0.1° resolution (approximately 8.5 km spatial resolution). This constitutes our input weather variables  $\mathcal{X}_{i,s}, i\in\mathcal{I},s\in\mathcal{T}$ used for modelling. This interpolation not only refines the resolution but, due to a cropping of the overall extent, ensures that the predictor and outcome variables are spatially aligned.

\begin{table}[h!]
    \centering
    \begin{tabular}{p{3cm} p{4.3cm} p{6cm}}
        \toprule
        \textbf{Source} & \textbf{Weather Variables} & \textbf{Description} \\
        \midrule

        ERA5 Dataset & 
        Air Temperature, Specific Humidity, Geopotential Height (500 hPa), Wind Velocity (U/V, 850 hPa), Total Column Water Vapour & 
        Historical weather state estimates from a numerical weather model with continuous observation assimilation. Provides 6-hourly records at 1-degree resolution. Used as inputs for our neural network models. \\
        \midrule

                E-OBS Rainfall Dataset & 
        Daily Total Precipitation & 
        Observational records collected from a network of weather stations. Provides daily records at 0.1-degree resolution. Used as the target dataset for our neural network models. \\
        \midrule

        IFS Precipitation Dataset for benchmarking & 
        Daily Total Precipitation & 
        Forecast precipitation estimates from the free-running IFS model. Unlike ERA5, these forecasts rely solely on internal model physics without continuous assimilation of precipitation observations. Used as a deterministic baseline for comparison. \\
        \bottomrule
    \end{tabular}
    \caption{Meteorological Datasets.}
    \label{tab:datasources}

\end{table}

\paragraph{IFS Benchmark data:}As a deterministic benchmark, we use precipitation predictions available from short-term forecast simulations with IFS as they were performed to provide the background state of the 4DVar data-assimilation when ERA5 was generated. 
There are two forecast simulations started each day at 6 am and 6 pm. We extract the precipitation fields for the first 12 hours of each simulation to reproduce daily precipitation - this is presently the optimal way to derive meaningful precipitation predictions from a dynamical model that is consistent with the large-scale fields in the ERA5 reanalysis data. Note that the IFS precipitation predictions provide only a deterministic prediction for each point in space and time, which requires simulating the full three-dimensional state of the global atmosphere, and are computationally expensive. In contrast, our mapping provides a probabilistically downscaled rainfall value, requiring only 6 local model fields as input, and can be generated at a much lower cost once the model is trained. Furthermore, the precipitation data was mapped directly from the native Gaussian grid of ERA5 (N320) to the high-resolution grid (\ang{0.1} longitude and latitude) of the observations and was not mapped to the coarse resolution grid at $\ang[parse-numbers=false]{{5}/{9}}$ longitude and $\ang[parse-numbers=false]{{5}/{6}}$ latitude first. The IFS precipitation predictions have therefore a higher spatial resolution ($\sim\SI{30}{\kilo\metre}$) when compared to the model fields that serve as input to our model, making it an especially challenging benchmark.

\subsection{Statistical Problem}
\label{sec:stat}

Our goal is to downscale rainfall over the UK by conditioning on weather predictors while portraying uncertainty and preserving spatial coherence. We are presented with predictors in the form of weather variables $\boldsymbol{\mathcal{X}}_{i,t}$ for days $t\in \mathcal{T}$ and location indices $i$ from the set $\mathcal{I}$ of locations considered. As we want to provide a probabilistic forecast at a future time $s\in\mathcal{T}$ for rainfall $\boldsymbol{\mathcal{Y}}_{s}=(y_{1,s},\ldots,y_{n,s})$ simultaneously over all locations in $\mathcal{I}$, we need to construct a conditional joint distribution of rainfall over the $n$ locations at time $s$, written $\mathbb{P}(\boldsymbol{\mathcal{Y}_{s}}|
\boldsymbol{\mathcal{H}}_s)$, conditioning on the history of predictor variables until time $s$ denoted as $ \boldsymbol{\mathcal{H}}_s \equiv \{ \boldsymbol{\mathcal{X}}_{i,t} : i\in \mathcal{I} \text{ and } t\leq s \}$. Further, we assume that this conditional dependency is $k$-Markovian in nature, that is  
$$\mathbb{P}(\boldsymbol{\mathcal{Y}_{s}}|
{\mathcal{X}}_{i,t} : i\in \mathcal{I} \text{ and } t\leq s) = \mathbb{P}(\boldsymbol{\mathcal{Y}_{s}}|
{\mathcal{X}}_{i,t} : i\in \mathcal{I} \text{ and } s-k\leq t\leq s).$$ For this work, we chose $k$ to be 28 days, meaning distributions of precipitation at a location only depend on the values of the model fields from the last 28 days, which is a reasonable assumption from a meteorological point of view.   

Since this conditional joint density is hard to estimate in one step, we decompose our approach into two sub-parts using Sklar's theorem \citep{sklar} on copulas, which are multivariate cumulative distribution functions (CDFs) for which the marginal probability distribution of each variable is uniform on the interval $[0, 1]$.

Sklar's theorem uses copula densities on $[0,1]^n$ to link the marginal density functions from $n$ dimensions together, thereby forming a joint probability density function (PDFs) for random variables in $\mathbb{R}^n$. But we notice that the observed rainfall $y_{i,t}$ is a non-negative random variable with a significantly high probability of zero, as a rainfall value for a dry day translates to a zero-valued observation, making our observations a case of \emph{zero-inflated data}, in which zero-valued data points are over-represented. 

Hence, to apply Sklar's theorem for modelling the joint distribution of rainfall, we consider our observations $\boldsymbol{\mathcal{Y}}_{s}$ being censored observations of a latent variable $\boldsymbol{\mathcal{Y}}_{s}^{*} \in \mathbb{R}^n$ where

\begin{equation*}
    \begin{aligned}
    y_{i,s}{=}\begin{cases}y_{i,s}^{*}, & \text{if $y_{i,s}^{*}>0$ }\\
                        0, & \text{if $y_{i,s}^{*} \leq0$}\end{cases}.
    \end{aligned}
\end{equation*}

We assume the conditional marginal densities of the latent variables $\boldsymbol{\mathcal{Y}}_{s}^{*}$ are $f_{i,s}^{*}(y_{i,s}^{*}|\boldsymbol{\mathcal{H}}_s) ,\, i\in\mathcal{I},\,s\in\mathcal{T}$ (with distributions  $F_{i,s}^{*}(y_{i,s}^{*}|\boldsymbol{\mathcal{H}}_s)$ ) at location $i \in \mathcal{I}$ and model the dependence between them with a unique copula $$\mathbf{c}^{*}\left(F_{1,s}^{*}(y_{1,s}^{*}|\boldsymbol{\mathcal{H}}_s),\ldots, F_{n,s}^{*}(y_{n,s}^{*}|\boldsymbol{\mathcal{H}}_s)|\boldsymbol{\mathcal{H}}_s, \mathcal{D}\right)$$ given $\boldsymbol{\mathcal{H}}_s$ and $\mathcal{D}$, the latter being a $n\times n$ matrix of spatial information for $n$ locations in $\mathcal{I}$. Hence Sklar's theorem \citep{sklar} provides us with a joint density, denoted $\pi(\boldsymbol{\mathcal{Y}}_{s}^{*}|
\boldsymbol{\mathcal{H}}_s)$, of the form 

\begin{equation*}  \pi^*\left(\boldsymbol{\mathcal{Y}}_{s}^{*}|\boldsymbol{\mathcal{H}}_s\right)=
   f_{1,s}^{*}\left(y^{*}_{1,s}|\boldsymbol{\mathcal{H}}_s\right) \cdot \ldots \cdot f^{*}_{n,s}\left(y^{*}_{n,s}|\boldsymbol{\mathcal{H}}_s\right)
   \cdot    \mathbf{c}^{*}\left(F^{*}_{1,s}\left(y^{*}_{1,s}|\boldsymbol{\mathcal{H}}_s\right), \ldots, F^{*}_{n,s}\left((y^{*}_{n,s}|\boldsymbol{\mathcal{H}}_s\right)|\boldsymbol{\mathcal{H}}_s, \mathcal{D}\right) .
\end{equation*}

We consider $f_{i,s}$ and $F_{i,s}$ are respectively the conditional marginal density and conditional marginal cumulative distribution function of rainfall at location $i$ and time $s$ given $\boldsymbol{\mathcal{H}}_s$, which corresponds to the probability density function (with respect to the Lebesgue $+ \delta_o$ measure)

\begin{equation*}
    f_{i,s}\left(y_{i,s}|\boldsymbol{\mathcal{H}}_s\right)
    =\begin{cases}
    f^{*}_{i,s}\left(y_{i,s}|\boldsymbol{\mathcal{H}}_s\right), & \text{if $y_{i,s}>0$ }\\
    F^{*}_{i,s}\left(0|\boldsymbol{\mathcal{H}}_s\right), & \text{if $y_{i,s} = 0$}
    \end{cases}
\end{equation*}

and the distribution function

\begin{equation*}
    F_{i,s}\left(y_{i,s}|\boldsymbol{\mathcal{H}}_s\right)
    =\begin{cases}
    F^{*}_{i,s}\left(y_{i,s}|\boldsymbol{\mathcal{H}}_s\right), & \text{if $y_{i,s}>0$ }\\
    p_{i,s} = F^{*}_{i,s}\left(0|\boldsymbol{\mathcal{H}}_s\right), & \text{if $y_{i,s} = 0$}
    \end{cases}.
\end{equation*}

In order to model the joint distribution of rainfall, our approach at first involves directly modelling the marginal densities $f_{i,s}$ (and correspondingly, distributions $F_{i,s}$) by leveraging established distributions of zero-inflated data. Examples of such distributions include a mixture model with mass at $\{0\}$ and a continuous parametric density for $y_{i,t}>0$ \citep{das1955fitting, shimizu1993bivariate}. To condition the marginal densities $f_{i,s}$ on $\boldsymbol{\mathcal{H}}_s$, we consider the parameters of this distribution family to be functions of $\boldsymbol{\mathcal{H}}_s$ in our novel approach termed the joint generalised neural model (JGNM). This model transforms  $\boldsymbol{\mathcal{H}}_s$ into informative summaries via a function parameterised by a neural network architecture. These summaries are then employed as predictors within a joint generalised linear model framework \citep{nelder1991generalized} to {jointly} model the different parameters of the marginal densities{, hence the name}. The initial transformation of predictors into summaries is designed to extract non-linear trends from the data. More specifically, the temporal and spatial dependency structures are captured through distinct neural network architectures, employing convolutional long short-term memory network \citep{ConvLSTMshi} modules. For each of the locations in $\mathcal{I}$ and times in $\mathcal{T}$, the JGNM is trained to output parameters which minimize the negative log-likelihood of rainfall under the parametric family model for $f_{i,s}(y_{i,s}|\boldsymbol{\mathcal{H}}_s)$. The detailed description of JGNM and the inference process are elaborated in Section~\ref{sec:marginal_deep}.

\begin{figure}[t]
    \centering
    \includegraphics[width=0.9\linewidth]{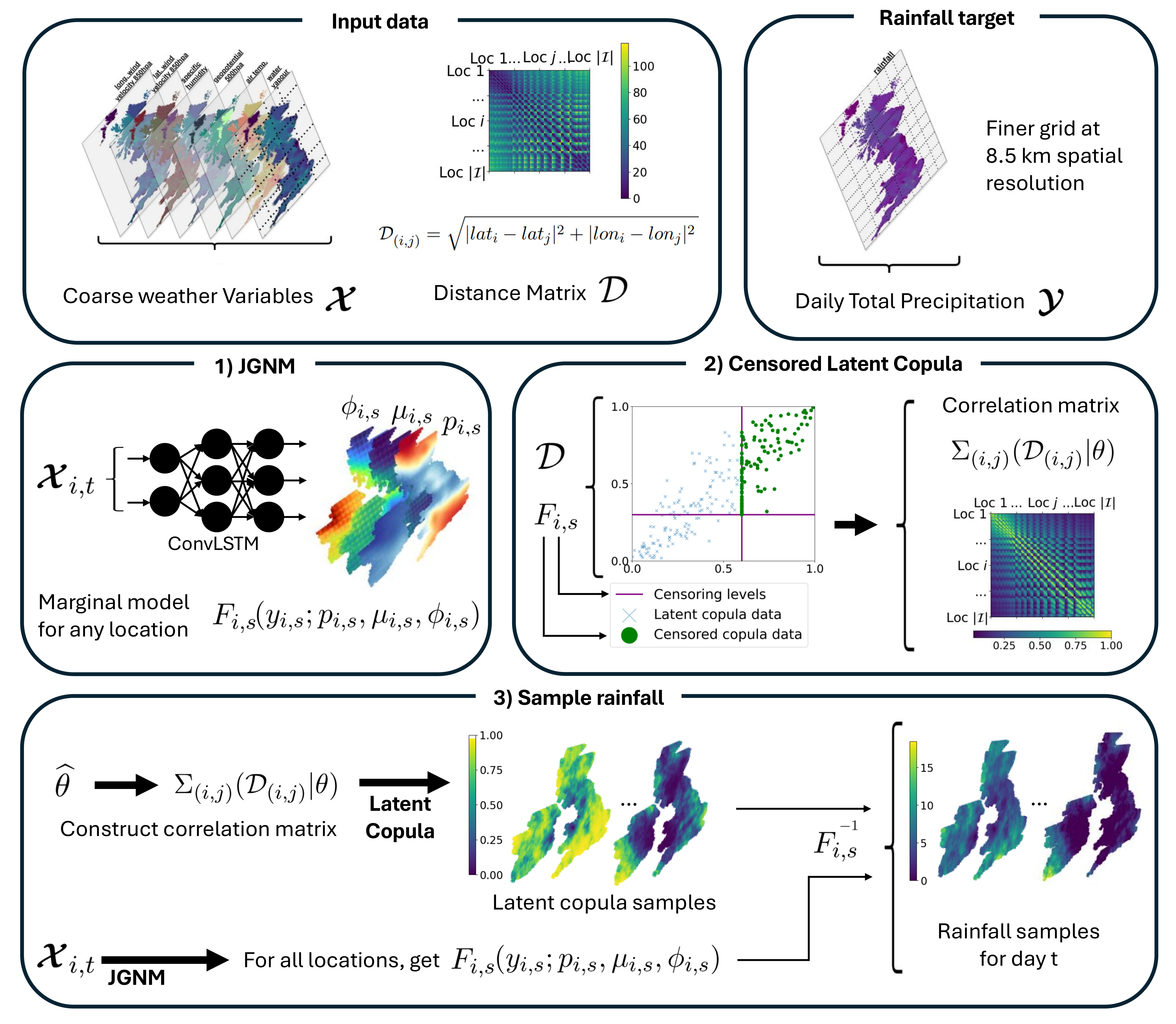}
    \caption{\textbf{Diagram of our Cens-JGNM model:} Our method takes weather variables and distances between locations as inputs with the goal of downscaling samples to higher resolution target rainfall. In part 1), we predict parameters for marginal distributions. In part 2), we model the dependence with a censored latent copula. Finally, in part 3), we produce rainfall samples by applying the inverse distributions to latent copula samples.}
    \label{fig:diagram}
\end{figure}

In the second part, to model the spatial dependency structure we assume a Gaussian copula for the latent continuous variables $\boldsymbol{\mathcal{Y}}_{s}^{*}$. To encapsulate the spatial dependency structure among the $n$ locations, we draw upon the literature on Gaussian processes \citep{rasmussen2006gaussian}. The {correlation} matrix $\Sigma$ of the Gaussian copula is determined by a covariance function $k$
, which offers a scalar measure of similarity between the spatial information of two points. This measure is often referred to as the kernel of the Gaussian process. The correlation matrix is generated by pairwise evaluation of the kernel on all points. For the sake of simplicity, we assume that the spatial dependency structure is time-independent, and hence $\Sigma$ does not depend on $\boldsymbol{\mathcal{H}}_s$. The specifics of the parametric form and the inference of the kernel parameter will be discussed in Section~\ref{sec:cens_copula}. It is important to note that the computation of the maximum likelihood estimator of the kernel parameters is infeasible as we do not have knowledge of $f_{i,s}^*(y_{i,s}^*)$ and $F_{i,s}^*(y_{i,s}^*)$ for $y_{i,s}^* \leq 0,\ \forall i \in \mathcal{I}.$ As a result, we propose an innovative estimation methodology based on the minimum Scoring Rules estimator \citep{dawid2016minimum, pacchiardi2022likelihood, alquier2022estimation}, which requires only simulations from the censored latent Gaussian copula. We provide a diagram of our model, termed Cens-JGNM, in Figure \ref{fig:diagram}.


\section{Modelling Marginals with the Joint Generalised Neural Model}
\label{sec:marginal_deep}

We will now describe the considered parametric model for the conditional marginal densities $f_{i,s}(y_{i,s}|\boldsymbol{\mathcal{H}}_s)$ and their corresponding distributions $F_{i,s}(y_{i,s}|\boldsymbol{\mathcal{H}}_s)$ in Section \ref{sec:JGNM}. Later in Section \ref{sec:JGNM_inference}, we describe how to train parameters of these densities with details of our model's neural architecture given in Section \ref{sec:JGNM_NN}.
 
\subsection{Joint Generalised Neural Model}
\label{sec:JGNM}
As our goal is to retain explainability in our modelling approach, for marginal densities, we opt for a parametric form. Additionally, this saves computational costs in Section~\ref{sec:cens_copula}, where repeated exact evaluations of the distributions are necessary. Due to the zero-inflated nature of rainfall data, we consider the following parametric form for the marginal density
\begin{equation*}
\begin{aligned}
    \mathcal{P}(y_{i,s} ; p,\mu_{i,s}, \phi_{i,s})
    =[1-p_{i,s}] \cdot\delta_{0}(y_{i,s})+p_{i,s} \cdot \mathcal{F}(y_{i,s};\mu_{i,s},\phi_{i,s}), 
\end{aligned}
\end{equation*}
$\text{for } y_{i,s} \in \mathbb{R}_{\geq 0},  i \in \mathcal{I}, s\in\mathcal{T}$, where $\delta_{0}(.)$ is the Dirac mass measure at 0, $p_{i,s} \in [0,1]$ is the probability of positive rainfall and with $\mu_{i,s}>0$ and $\phi_{i,s}>0$ symbolising the mean and dispersion parameters\footnote{Note here that $\phi_{i,s}$ refers to the dispersion parameter while $\phi$ without subscripts refers to the Gaussian PDF.} of $\mathcal{F}$ which is a continuous density defined on $\mathbb{R}_{>0}$.  Existing studies \citep{das1955fitting,fealy2007statistical,paper10,ailliot2015stochastic,holsclaw2017bayesian,bertolacci2019climate,xie2023assessment} argue in favour of a gamma distribution as the choice of $\mathcal{F}$, following which we will consider the zero-gamma mixture in the rest of this work  with density given by
\begin{equation*}
\begin{aligned}
    \mathcal{P}(y_{i,s} ; p,\mu_{i,s}, \phi_{i,s})
    =
    \left[1-p_{i,s}\right] \cdot\delta_{0}\left(y_{i,s}\right)+p_{i,s} \cdot\left(\frac{y_{i,s}}{{\phi_{i,s}} \cdot \mu_{i,s}}\right)^{1 / \phi_{i,s}} \frac{1}{y_{i,s}} \exp \left(-\frac{y_{i,s}}{\phi_{i,s} \cdot\mu_{i,s}}\right) \frac{1}{\Gamma(1 / \phi_{i,s})},
\end{aligned}
\end{equation*}
$\text{ for } y_{i,s} \in \mathbb{R}_{\geq 0},  i \in \mathcal{I}, s\in\mathcal{T}$. We further assume that the marginal distributions are influenced by the conditioning predictor variables $(\boldsymbol{\mathcal{H}}_s)$ exclusively through their parameters.\par
For any given $y_{i,s}$ and a relevant subset of predictors $\boldsymbol{x}_{i,s} \subseteq\boldsymbol{\mathcal{H}}_s$, following the literature on generalised linear models (GLMs) \citep{Dunn_glm} and joint generalised linear model (JGLMs) \citep{nelder1991generalized}, we can model the parameters $(p_{i,s},\mu_{i,s},\phi_{i,s})$ specific to location $i$ and time $t$ as,  
\begin{equation}
\label{eq:jGLM}
    \begin{aligned}
g_p(p_{i,s})=\alpha_{0} +\left\langle\boldsymbol{\alpha}, \boldsymbol{x}_{i,s}\right\rangle \\
g_{\mu}(\mu_{i,s})=\beta_{0}+\left\langle\boldsymbol{\beta}, \boldsymbol{x}_{i,s}\right\rangle \\
g_{\phi}(\phi_{i,s})=\gamma_{0}+\left\langle\boldsymbol{\gamma}, \boldsymbol{x}_{i,s}\right\rangle 
    \end{aligned}
\end{equation}
where $(\alpha_{0},\boldsymbol{\alpha}),(\beta_{0},\boldsymbol{\beta})$ and $(\gamma_{0},\boldsymbol{\gamma})$ are regression coefficients and $g_{p_{i,s}}(p_{i,s})= \log (p_{i,s} /(1-p_{i,s})), p_{i,s}\in (0,1) $, $g_{\mu_{i,s}}(\mu_{i,s})= \log(\mu_{i,s})$, $g_{\phi_{i,s}}(\phi_{i,s})= \log(\phi_{i,s})$ are the link functions. In practice, since we do not require reversibility of our link function, we are free to choose an inverse link function that has an appropriate range, but provides a more stable set of gradients during training. Table ~\ref{tab:mean_function_gamma} details the specific inverse link function used in our work.

While a JGLM would allow us to have location and time-dependent models for marginal densities $f_i(y_{i,s}|\boldsymbol{\mathcal{H}}_s)$, we can still seek improvement in the complexity of the model. Notably, the relationship between predictors and density parameters is linear in the coefficients. As such, non-linear relationships in the modelling of parameters will not be taken into account by the marginals. To remedy this, a further improvement of GLMs can be considered by incorporating neural networks \citep{Goodfellow}.

Similar to the DeepGLM model introduced in \cite{DGLM}, we refine the predictors $\boldsymbol{x}_{i,s} \subseteq\boldsymbol{\mathcal{H}}_s$ by training a neural network $\mathcal{N}_{\psi}$ parameterised by $\psi$ to obtain refined features $\mathcal{N}_{\psi}(\boldsymbol{x}_{i,s})$. These refined features are used in the place of $\boldsymbol{x}_{i,s}$ as predictor variables in the Equation \eqref{eq:jGLM} to arrive at our joint generalised neural model, 
\begin{equation*}
    \begin{aligned}
        g_p(p_{i,s})= \alpha_{0}+\left\langle\boldsymbol{\alpha}, \mathcal{N}_{\psi}(\boldsymbol{x}_{i,s})\right\rangle \\
        g_{\mu}(\mu_{i,s})=\beta_{0}+\left\langle\boldsymbol{\beta}_i, \mathcal{N}_{\psi}(\boldsymbol{x}_{i,s})\right\rangle \\
        g_{\phi}(\phi_{i,s})=\gamma_{0}+\left\langle\boldsymbol{\gamma}, \mathcal{N}_{\psi}(\boldsymbol{x}_{i,s})\right\rangle 
    \end{aligned}
\end{equation*}
where $(\alpha_{0},\boldsymbol{\alpha}),(\beta_{0},\boldsymbol{\beta})$ and $(\gamma_{0},\boldsymbol{\gamma})$ are regression coefficients in $\mathbb{R}$ and $g_p(.),g_{\mu}(.),g_{\phi}(.)$ represent the link functions.

Thus, our JGNM model can be seen as a distributional learner capable of outputting multiple parameters given initial weather predictors by capturing linear as well as non-linear relationships present in the data. {Unlike models that output a field of predictions over fixed
 points (e.g. Harris et al. (2022); Vosper et al. (2023)), our model only needs predictors to
 output parameterised densities. Thus, given predictors $\boldsymbol{x}_{i,s}$for a new location $i$ expected to follow
 the same precipitation patterns as in $\mathcal{I}$, our model can be applied to such locations outside the training
 set, thereby enabling, for example, principled interpolation.} Indeed, one only needs to input initial predictors $\boldsymbol{x}_{i,s}$ into $\mathcal{N}_{\psi}$ in order to have the JGNM parameterise a distribution; no further distinction is made between locations given the input, meaning additional locations can be included into our model without needing to retrain it.\par

\subsection{Inference for the JGNM}
\label{sec:JGNM_inference}
To ensure our model aligns with the selected distributional assumptions, we learn the aggregated parameters $\left(\psi,\alpha_0,\boldsymbol{\alpha},\beta_0,\boldsymbol{\beta},\gamma_0,\boldsymbol{\gamma}  \right)$ using the likelihood specific to the assumed marginal model in the loss function of the JGNM. 

For the gamma mixture, the JGNM loss function contains two segments. The first segment pertains to $p_{i,s}$ and adopts the form of a logistic loss. The second segment is associated with the parameters $\mu_{i,s},\phi_{i,s}$ and is represented by the negative log-likelihood (NLL) of the corresponding gamma density. In instances of zero rainfall, only the logistic component is considered, while in the event of rainfall, both components are evaluated.\par
The component in the JGNM loss function accounting for rainfall occurrence in terms of $p_{i,s}$ is the logistic loss, also known as binary cross entropy, between occurrence and $p_{i,s}$, written as
\begin{equation*}
   \mathcal{L}(p_{i,s}|y_{i,s}) = -\mathbb{I}\{y_{i,s}=0\}\log(p_{i,s})-(1-\mathbb{I}\{y_{i,s}=0\})\log(1-p_{i,s}).
\end{equation*}
Under a GLM parameterisation the Gamma density for positive rain $y_{i,s}>0$ is written as
\begin{equation*}
\begin{aligned}
    f_{i}(y_{i,s}|\boldsymbol{\mathcal{H}}_s)&=\mathcal{P}(y_{i,s} ; p_{i,s},\mu_{i,s}, \phi_{i,s})\\
    &=\left(\frac{y_{i,s}}{\phi_{i,s} \mu_{i,s}}\right)^{1 / \phi_{i,s}} \frac{1}{y_i} \exp \left(-\frac{y_i}{\phi_{i,s} \mu_{i,s}}\right) \frac{1}{\Gamma(1 / \phi_{i,s})}
\end{aligned}
\end{equation*}

leading to a negative log-likelihood corresponding to the loss for positive rain  of the form
\begin{equation*}
    \mathcal{L}(\mu_{i,s},\phi_{i,s}|y_{i,s})=\left(\frac{1}{\phi_{i,s}}-1\right)\log(y_{i,s})-\frac{\log(\phi_{i,s}\mu_{i,s})}{\phi_{i,s}}-\frac{y_{i,s}}{\mu_{i,s}\phi_{i,s}}-\log(\Gamma(\phi_{i,s}^{-1})).
\end{equation*}\par

\subsection{Network Architecture}
\label{sec:JGNM_NN}
The neural network component [$\mathcal{N}_{\psi}$] of the JGNM employs a unique architecture to capture the spatio-temporal properties of the data. Our proposed model contains a convolutional long short-term memory network (ConvLSTM) \citep{ConvLSTMshi} — a neural module that combines convolutional Neural Networks (CNN) and Long Short-Term Memory (LSTM) — to extract both spatial and temporal trends effectively. ConvLSTM leverages the strengths of LSTM models, which are recursive models appropriate for time-series data, and convolutional networks, a notable architecture for 2D/3D image data. A comprehensive description of the architecture can be found in Appendix~\ref{apdx:JGNM_NN}.

In addition to the standard ConvLSTM, our architecture is novel in its integration of several key modules that differentiate it from conventional approaches. Notably, the Time Distributed 2D Convolution Layer (TD2L) ensures uniform feature extraction across time steps, while the Time Distributed Temporal Downscaling layer (TDTD) employs a convolutional self-attention mechanism to dynamically aggregate temporal information. Furthermore, the Time Distributed Mean Function Layer, composed of dedicated single-layer perceptrons, provides tailored inverse link functions for each parameter of the predictive distribution. Instead of employing simple temporal aggregation, we introduce a Convolutional Self-Attention layer to facilitate the transition within our model from 6-hourly weather field data to a 24-hourly representation.

Overall, our network comprises several key components with specific sizes and layer counts. The initial TD2L uses one 2D convolution to transform \(16 \times 16 \times 6\) input patches into \(16 \times 16 \times 64\) feature maps. This is followed by two bi-directional ConvLSTM layers with skip connections to robustly capture spatio-temporal dependencies. The TDTD layer then groups every four consecutive time steps, employing a CSA module that uses 4 attention heads (each with a hidden dimension of 64) to compute dynamic, weighted feature aggregations. This multi-head CSA module enhances the model’s ability to integrate information across time before the final downscaling. Finally, the Time Distributed Mean Function Layer uses three perceptrons to produce the parameters of our predictive distribution.


\section{Censored Latent Gaussian Copula}
\label{sec:cens_copula}

In this section, we centre our attention on the dependency structure model described in Section \ref{sec:cens_copula_spadep} and the associated challenges in its estimation, as discussed in Section \ref{sec:cens_copula_mle}. Our proposed solution is a novel estimation approach based on minimum scoring rules detailed in Section \ref{sec:cens_copula_minSR}, which we validate on simulated data in Section \ref{sec:cens_copula_simdata}.\par
For the latent variables $\boldsymbol{\mathcal{Y}}_{s}^{*}$, we assume that their dependence can be modelled by a Gaussian copula $\mathbf{c}$, with density
\begin{equation*}
    \begin{aligned}
    &\mathbf{c}^{*}\left(F^{*}_{1,s}\left(y^{*}_{1,s}|\boldsymbol{\mathcal{H}}_s\right), \ldots, F^{*}_{n,s}\left((y^{*}_{n,s}|\boldsymbol{\mathcal{H}}_s\right)|\boldsymbol{\mathcal{H}}_s, \mathcal{D}\right)  \\  
         &= \mathbf{c}(\Phi(z_{1,s}^{*}), \ldots , \Phi(z_{n,s}^{*})|\Sigma(\mathcal{D}|\theta))
         =
    \frac{\phi_n(z_{1,s}^{*},\ldots, z_{n,s}^{*}|\mathbf{0},\Sigma(\mathcal{D}|\theta)) }{\prod_{i=1}^{n}\phi(z_{i,s}^{*})}
    \end{aligned}
\end{equation*}
where $z_{i,s}^{*}=\Phi^{-1}\left(F_{i,s}^{*}(y_{i,s}^{*}|\boldsymbol{\mathcal{H}}_s)\right)$, $\phi$ and $\Phi$ are the PDF and CDF of univariate Gaussian distributions, and $\phi_n$ a Gaussian multivariate density with mean $\mathbf{0}$ and correlation matrix $\Sigma(\mathcal{D}|\theta)$ which depends on a matrix of spatial information $\mathcal{D}$ through a parameter $\theta$. {We construct $\mathcal{D}$ as the Euclidean distance based on the latitude and longitude of locations, the details of which are provided in Appendix~\ref{app:dist_const} where we also investigate and decide against using topographical information for $\mathcal{D}$.} We assume the dependence structure imposed by this correlation to be constant in time, hence dropping the dependence on $\boldsymbol{\mathcal{H}}_s$ in the construction of $\Sigma$. 
{This is a tradeoff of simplicity for modelling purposes and sophistication for predictive prowess.}
In this way, we model the joint \emph{relative} intensity as a function of space, while the exact amounts also depend on temporal information through the marginals.
Due to the Gaussianity assumption on $\mathbf{c}$, the only object to infer is the correlation matrix which incorporates the spatial dependence into our model.

\subsection{Modelling of Spatial Dependency}
\label{sec:cens_copula_spadep}
This reliance on $\mathcal{D}$ when constructing the latent Gaussian copula $\mathbf{c}$ induces the spatial dependence on latent variables $\boldsymbol{\mathcal{Y}}_s^{*}$ and hence on the observations $\boldsymbol{\mathcal{Y}}_s$.
To construct the correlation, we reference the existing literature on Gaussian processes \citep{rasmussen2006gaussian} and apply a Gaussian process kernel as an element-wise transformation of the initial matrix $\mathcal{D}$, which contains spatial information. This transformation ensures the resulting matrix's positive definiteness and thus its validity as a covariance of a Gaussian distribution. The choice of kernel $k(\mathcal{D}|\theta)$ specifies this transformation and typically depends on a few parameters, here denoted by $\theta$. We use the Matérn kernel \citep{matern2013spatial}, a common choice in spatial statistics \citep{gerber2021fast,richards2024neural}, which provides us with a correlation matrix $\Sigma(\mathcal{D}|\theta)$ of the following form:
\begin{equation*}
    \Sigma_{(i,j)}(\mathcal{D}_{(i,j)}|\theta) =  k(\mathcal{D}_{(i,j)}| \theta) = \frac{2^{1-\nu}}{\Gamma(\nu)}\cdot\left(\sqrt{2 \nu} \frac{\mathcal{D}_{(i,j)}}{\theta}\right)^{\nu} \cdot K_{\nu}\left(\sqrt{2 \nu} \frac{\mathcal{D}_{(i,j)}}{\theta}\right)
\end{equation*}
where $\Gamma$ is the gamma function and $K_{\nu}$ is the modified Basel function of the second kind, noting that the expression depends on the unknown lengthscale $\theta\, (\in \mathbb{R}_{>0})$  and additional parameter $\nu$ usually chosen by the user. In this work, we would consider $\nu$ to be fixed at 0.5 {as for values of $\nu=n+1/2, n\in\mathbb{N}$ the Matérn kernel has an equivalent simpler expression (as an exponential multiplied by a polynomial of order $n$), granting us a computational speedup \citep{guttorp2006studies}} (see details in Appendix \ref{apdx:optim}). \par
A convenient consequence of such a construction for the correlation matrix is that it becomes straightforward to include additional locations (if we expect them to follow the same pattern as our training set) in our approach even after fitting it to data. Indeed, as long as one has relevant spatial information about a new location, the initial distance matrix $\mathcal{D}$ can be expanded to include that new location by applying the learned kernel to the distances, including it in the dependence structure.

\subsection{Infeasibility of Maximum Likelihood Estimation}
\label{sec:cens_copula_mle}
Traditionally, inference for a Gaussian copula is done using maximum likelihood estimation (MLE) on its density. However, our Gaussian copula is latent, meaning we do not have access to direct realisations $z_{i,s}^{*}=\Phi^{-1}\left(F_{i,s}^{*}(y_{i,s}^{*}|\boldsymbol{\mathcal{H}}_s)\right)$ assumed to come from it. Instead, we examine the joint density of observed data conditional on the history $\pi\left(\boldsymbol{\mathcal{Y}}_{s}|\boldsymbol{\mathcal{H}}_s\right)$. {It can be obtained by considering the joint density of the latent variables $\pi^*\left(\boldsymbol{\mathcal{Y}}^{*}_{s}|\boldsymbol{\mathcal{H}}_s\right)$ and marginalising over censored locations denoted as $\mathcal{C}_s=\{i\in\mathcal{I}:y_{i,s}=0\}$. The marginalisation is done up to the censoring thresholds of $0$, to give the marginalised mass under the latent joint density $\pi^*\left(.|\boldsymbol{\mathcal{H}}_s\right)$ to censored events under $\pi\left(.|\boldsymbol{\mathcal{H}}_s\right)$ in the form of point mass. This is formalised as:}
\begin{eqnarray*} 
\label{eq:unav_mle}
\pi\left(\boldsymbol{\mathcal{Y}}_{s}|\boldsymbol{\mathcal{H}}_s\right)
&=& \int^{0}_{-\infty} \ldots \int^{0}_{-\infty}
\pi^*\left(\boldsymbol{\mathcal{Y}}_{s}^{*}|\boldsymbol{\mathcal{H}}_s\right) \prod_{i\in\mathcal{C}_s} dy^{*}_{i,s} \\
&=& \int^{0}_{-\infty} \ldots \int^{0}_{-\infty}f_{1,s}^{*}\left(y^{*}_{1,s}|\boldsymbol{\mathcal{H}}_s\right) \cdot \ldots \cdot f^{*}_{n,s}\left(y^{*}_{n,s}|\boldsymbol{\mathcal{H}}_s\right)\\
   &&\cdot    \mathbf{c}^{*}\left(F^{*}_{1,s}\left(y^{*}_{1,s}|\boldsymbol{\mathcal{H}}_s\right), \ldots, F^{*}_{n,s}\left((y^{*}_{n,s}|\boldsymbol{\mathcal{H}}_s\right)|\mathcal{D},\theta\right) \prod_{i\in\mathcal{C}_s} dy^{*}_{i,s}\\
&=&  \int^{0}_{-\infty} \ldots \int^{0}_{-\infty}f_{1,s}^{*}\left(y^{*}_{1,s}|\boldsymbol{\mathcal{H}}_s\right) \cdot \ldots \cdot f^{*}_{n,s}\left(y^{*}_{n,s}|\boldsymbol{\mathcal{H}}_s\right)
\\&&
\cdot\frac{\phi_n\left(z_{1,s}^{*},\ldots,z_{n,s}^{*}|\mathbf{0},\Sigma(\mathcal{D}|\theta)\right)}
        {\prod_{i=1}^{n}\phi(z_{i,s}^{*})}\prod_{i\in\mathcal{C}_s} dy^{*}_{i,s}.
\end{eqnarray*}
{While we know the form of the copula by assumption, performing MLE on this expression is in-feasible due to the unavailability of marginals $f_{i,s}^*, F_{i,s}^*$ over $(-\infty,0]$ which are needed for the integration over censored values.}  Thus, we cannot rely on MLE for inference of the kernel parameter $\theta$. 

To circumvent the problematic intractable likelihood expression, we develop a methodology using minimum scoring rules estimation \citep{dawid2016minimum} for the inference of parameters of Gaussian copulas, suitable to our case of censored latent Gaussian copulas.\par

\subsection{Minimum Scoring Rule Censored Gaussian Copula}
\label{sec:cens_copula_minSR}

Introduced in \cite{dawid2016minimum, JMLR:v23:21-0061},  the idea of minimum scoring rule estimation is to simulate realisations from the model for a chosen parameter $\theta \in \Theta$ and select the best such parameter by comparing the simulated data against observations. This comparison is done using scoring rules \citep{doi:10.1198/016214506000001437} which under suitable conditions define a divergence measure between two distributions and can be generalised to compare two data sets. As defined in \cite{doi:10.1198/016214506000001437}, a scoring rule $S(\mathcal{P}_{\theta},\mathbf{z})$ is a function between a distribution $\mathcal{P}_{\theta}$ and observed data $\mathbf{z}$ as a realisation of a random variable $\mathbf{Z}\sim \mathcal{P}^{*}$. Hence a minimum scoring rule estimate becomes, 
\begin{equation*}
    \widehat{\theta} = \argmin_{\theta \in \Theta}S\left(\mathcal{P}_{\theta},\mathbf{z}\right).
\end{equation*}
If the scoring rule is strictly proper (as defined in Appendix~\ref{app:SR_div}), this can be shown as asymptotically (over repeated samples $\mathbf{z}$) equivalent to minimizing a statistical divergence between $\mathcal{P}_{\theta}$ and $\mathcal{P}^{*}$. Considering the negative log probability density function of $\mathcal{P}_{\theta}$ evaluated at the observation as a score function, we can arrive at the maximum likelihood estimation scheme and correspondingly to the Kullback-Liebler divergence. We provide more details regarding these connections in Apprendix~\ref{app:SR_div}. Due to the intractability of the probability density function of the latent Gaussian copula, we choose the energy score,
\begin{equation*}
\label{eq:ensr}
S_E^{(\beta)}(P, \mathbf{z}) = 2 \cdot \mathbb{E} \left[\| \mathbf{Z} - \mathbf{z}\|_2^\beta\right] - \mathbb{E}\left[\|\mathbf{Z} - \mathbf{Z}'\|_2^\beta\right] ,\quad  \mathbf{Z} \independent \mathbf{Z}' \sim P,
\end{equation*}
which is a strictly proper scoring rule for $ \beta \in (0,2) $. In our study, we fix $\beta=0.5$ {for better optimisation stability based on preliminary experiments}, which we will assume henceforth for all applications of the energy score. If we can draw $m$ identical and independent samples $\mathbf{z}'=\left(\mathbf{z}_{1}^{\prime}, \ldots, \mathbf{z}_{m}^{\prime}\right)$ from the distribution $\mathcal{P}_\theta$, then an unbiased estimate of the energy score $S_E^{(\beta)}(\mathcal{P}_\theta,\mathbf{z})$ can be constructed as follows:
\begin{equation*}
\label{eq:ue_ensr}
\widehat{S}_E^{(\beta)}(\mathbf{z}', \mathbf{z}) = \frac{2}{m} \sum_{j=1}^m\left\|{\mathbf{z}_{j}^{\prime}}-\mathbf{z}\right\|_2^\beta-\frac{1}{m(m-1)} \sum_{\substack{ k=1 \\ k \neq j}}^m\left\|{\mathbf{z}_{j}^{\prime}}-{\mathbf{z}_{k}^{\prime}}\right\|_2^\beta.
\end{equation*}

The inference of the parameter for the censored latent Gaussian copula model depends on the crucial observation that we can simulate from this model for any given parameter $\theta$. As a Gaussian copula is associated to a Gaussian with a given correlation matrix $\Sigma$ \citep{nelsen2006introduction}, we can infer the correlation matrix directly on the Gaussian scale by transforming the observed data to the Gaussian scale as $z_{i,s} = \Phi^{-1}(F_{i,s}(y_{i,s}|\boldsymbol{\mathcal{H}}_s)) \in \mathbb{R}$. For a censored Gaussian copula, we correspondingly deal with a censored Gaussian density. To compute the minimum scoring rule estimate of $\theta$, we define simulations $\mathbf{z}_s^{\prime}=(z_{1,s}^{\prime},\ldots,z_{n,s}^{\prime})$ from the censored Gaussian density (corresponding to the censored Gaussian copula) as
\begin{equation}
\label{eq:censoring_Gaussian_scale}
    z_{i,s}^{\prime}
    =\begin{cases}
    z^{*}_{i,s}, & \text{if $z^{*}_{i,s}>d_{i,s}$ }\\
    d_{i,s}, & \text{if $z^{*}_{i,s}\leq d_{i,s}$}
    \end{cases}
\end{equation}
where $\mathbf{z}^{*}_s=(z^{*}_{1,s},\ldots,z^{*}_{n,s})$ is a simulation from the uncensored Gaussian density $\mathcal{N}( \cdot |\mathbf{0},\Sigma(\mathcal{D}|\theta))$, $d_{i,s}=\Phi^{-1}(1-p_{i,s})$ and $p_{i,s}$ corresponds to the mass at 0 for the marginal density model $F_{i,s}(\cdot|\boldsymbol{\mathcal{H}}_s)$.
Further, we also transform observed rainfall $\left(y_{i,s}: 1\leq i \leq n\right)$ to $\mathbf{z}_s=\left(z_{i,s}: 1\leq i \leq n\right)$ which is the corresponding output of the (assumedly true) censored Gaussian. We do so by applying the inferred marginal distribution $F_{i,s}(\cdot|\boldsymbol{\mathcal{H}}_s)$ to observation $y_{i,s}$, getting $z_{i,s} = \Phi^{-1}(F_{i,s}(y_{i,s}|\boldsymbol{\mathcal{H}}_s))$ at each location $i$ and time $s$. Comparing the transformed observations $\mathbf{z}_s$ with $m$ samples from the censored Gaussian with correlation parameterised by $\theta$, denoted by  $\mathbf{Z}'_{\theta,s}=\left\{{\mathbf{z}_{s}^{\prime}}^{(j)}=(z_{1,s}^{\prime},\ldots,z_{n,s}^{\prime})^{(j)} \right\}_{j=1}^m$, for each $s \in \mathcal{S}$, we can compute the minimum scoring rule estimate as follows:
\begin{eqnarray*}
\label{eq:minSRest}
    \widehat{\theta} 
    =\argmin_{\theta \in \Theta} \sum_{s \in \mathcal{S}}\hat{S}^{(\beta)}_{\mathrm{E}}\left(\mathbf{Z}'_{\theta,s},\mathbf{z}_s\right).
\end{eqnarray*}

\begin{figure}
\centering
\begin{subfigure}[t]{0.5\textwidth}
    \includegraphics[width=\textwidth,valign=t]{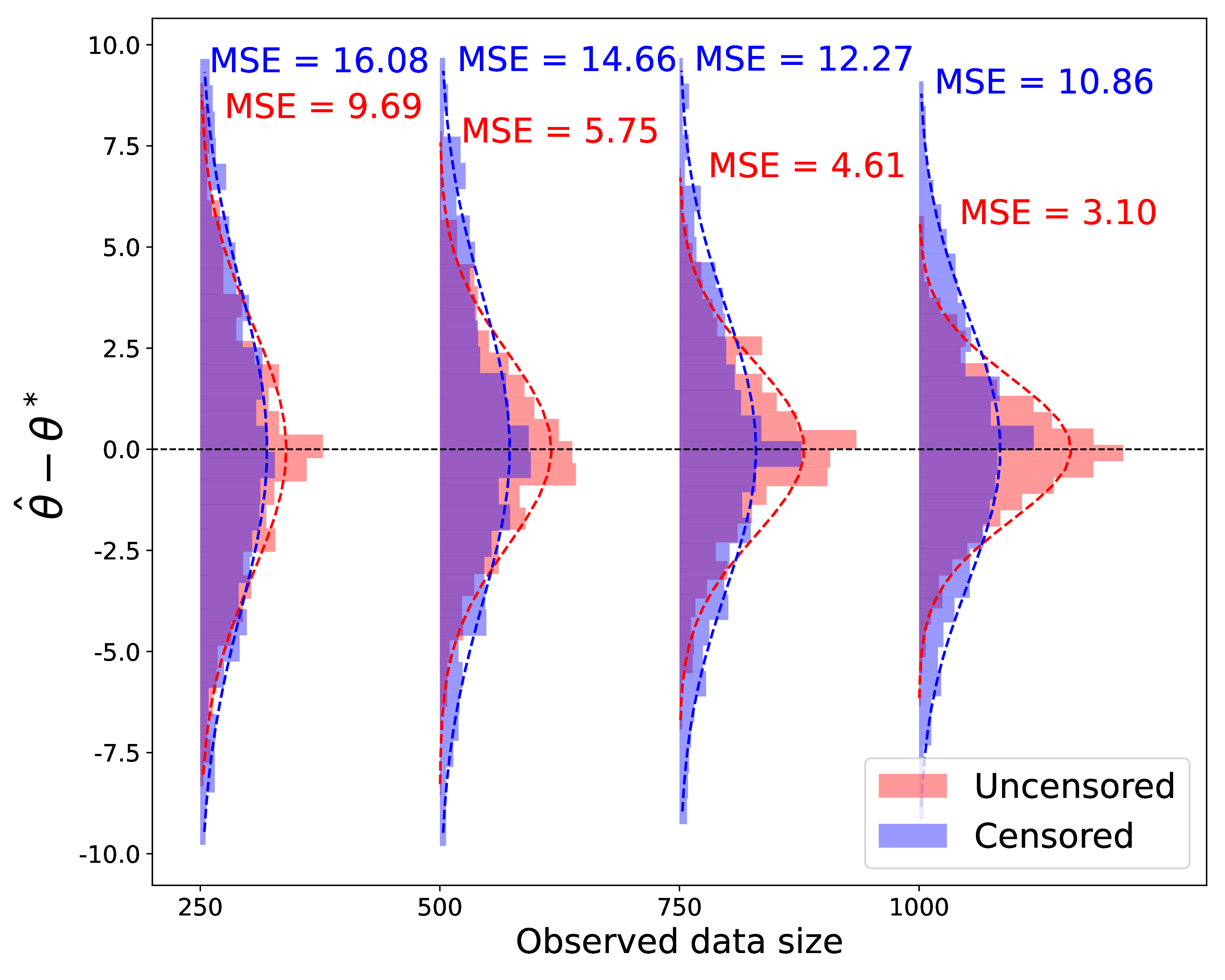}
    \caption{Parameter estimation error}
    \label{fig:subfigure_a_SR_test}
\end{subfigure}%
\begin{subfigure}[t]{0.5\textwidth}
    \includegraphics[width=\textwidth,valign=t]{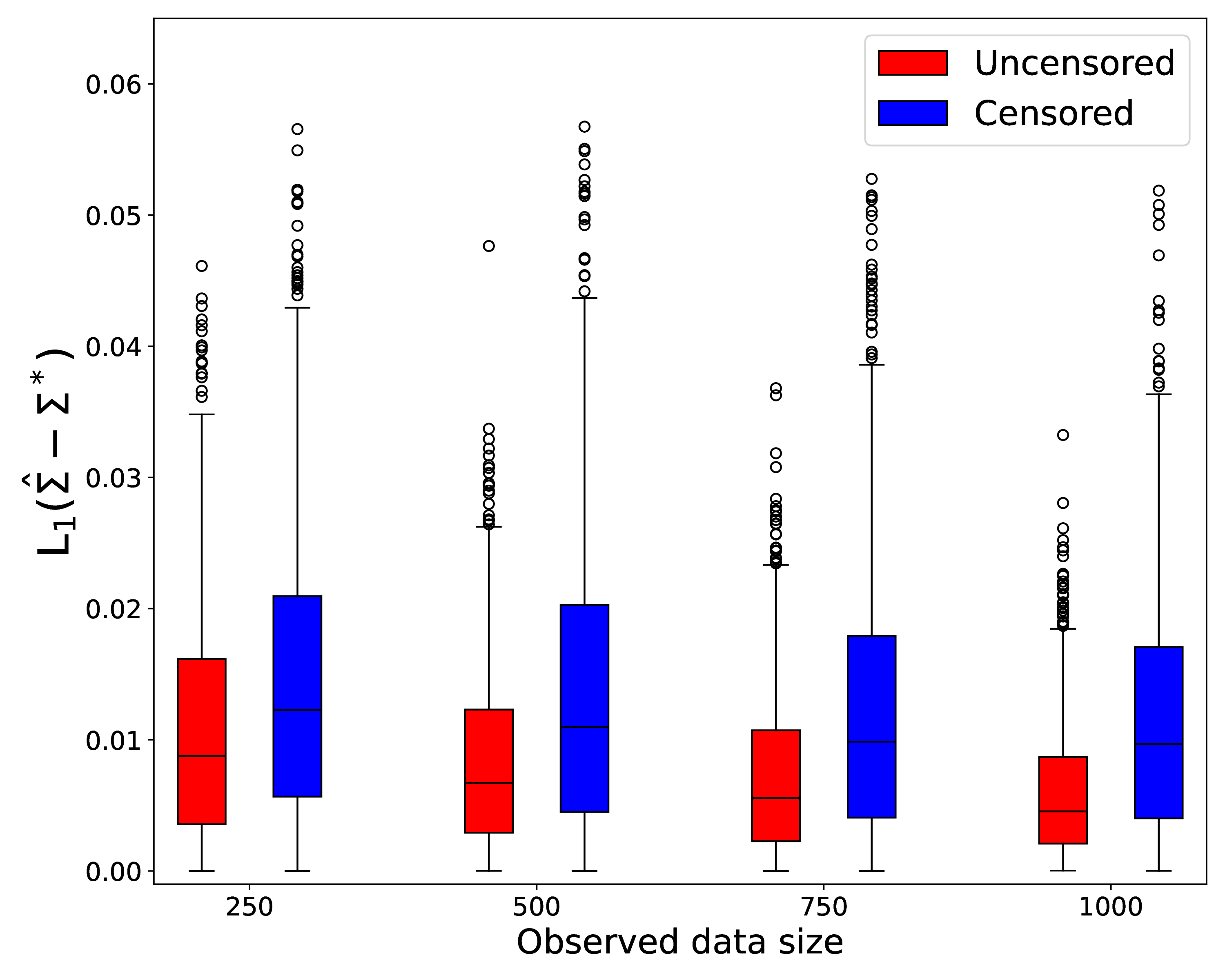}
    \caption{L$_1$ norm of correlation matrix}
    \label{fig:subfigure_b_SR_test}
\end{subfigure}

\caption{\textbf{Simulation study:} We fit a regular and a censored copula for 1000 replicates with our method, repeating this for increasing sample sizes. (a) The parameter estimation error distributions for both cases with the associated MSE, with the dotted lines representing a Gaussian approximation of the error density. (b) The L$_1$ norm compares the true and estimated matrix obtained with the fitted parameter $\hat{\theta}$. Both the censored and uncensored copulas show a decreasing error with the observed data size, occurring faster for the latter.}
\end{figure}

\subsection{Inference on simulated data}
\label{sec:cens_copula_simdata}

{To empirically validate our estimation procedure, we consider a simulation study with known parameter values. The experimental setup is as follows. For a given observed sample size of $\mathcal{T}=250,500,750,1000$ we sample from a spatial Gaussian copula with a correlation matrix $\Sigma^*$ parameterised using $\theta^*=35$ as described in Section \ref{sec:cens_copula_spadep}, and from a censored Gaussian copula with identical correlation where the censoring levels are $\{1-p_i\}_{i=1}^{|\mathcal{I}|} \sim U[0.5,0.95]$. We fix $|\mathcal{I}|=3$ since under a correctly specified model the correlation kernel is applied entry-wise, hence rendering all dimension pairs informative about $\theta$, meaning inference can be conducted with any subset of locations. We optimise $\theta$ using our Minimum Scoring Rule approach of Section \ref{sec:cens_copula_minSR} with the gradient-based Adam \citep{kingma2014adam} optimiser to convergence to obtain our parameter estimate $\hat{\theta}$, in turn getting $\hat{\Sigma}$. The size of simulations for the Scoring Rule is always set to the observation sample size. We repeat this experiment 1000 times with different optimisation initialisations in $U[30,40]$ to imitate our initialisation in practice, see Appendix \ref{apdx:optim}.}\par
{In Figure \ref{fig:subfigure_a_SR_test}, we show the distribution of estimation errors for $\theta$ with sample size as well as the mean squared error, in blue and red respectively for the censored and uncensored Gaussian copulas. Our minimal SR inference approach recovers the true parameter for both copula types, with the estimation error shrinking with sample size and approximately following a Gaussian density (shown as dotted lines fitted on the errors) centred at the true parameter value. In Figure \ref{fig:subfigure_b_SR_test}, we show the L$_1$ norm (see Appendix \ref{apdx:diagnostics_L1}) of correlation matrix differences ${\Sigma_{\hat{\theta}}}-\Sigma_{\theta^*}$, depicting a similar pattern, where the estimated correlation matrix matches the true matrix better with increasing sample sizes.}\par
{In the uncensored case, the error variance decreases more rapidly than in the censored case. This reflects the loss of information due to censoring. Practically, instead of observing the full range of copula samples $[0,1]^{n}$, we only observe the range $[1-p_{1,s},1]\times\ldots\times[1-p_{n,s},1]$, making only a fraction of the samples fully observed, effectively diminishing our sample size. While the censored data still carries information about the correlation structure, there is a loss of information, which explains the slower rate of decreasing error in Figure~\ref{fig:subfigure_b_SR_test}.


\section{Real Data Scenario}
\label{sec:real_data}
Finally, we apply our approach to the data set presented in Section \ref{sec:data_stat}. We conduct a comprehensive evaluation by comparing our proposed method with other benchmark probabilistic downscaling methods using standard diagnostics of probabilistic prediction [Appendix~\ref{apdx:diagnostics}] in Section~\ref{sec:comp_benchmark}, in addition to a deterministic precipitation prediction provided by IFS data. We then assess the spatial coherence of the generated samples from all these methods in Section~\ref{sec:spat_coh}. Finally, we investigate the effect of varying lengths of training data on the performance of our approach in Section~\ref{sec:len_traindat}. 

\subsection{Comparison with benchmark models}
\label{sec:comp_benchmark}
We begin by comparing our method (denoted Cens-JGNM) to two competing benchmark methods: the variational auto-encoder generative adversarial network (VAE-GAN) \citep{Harris2022} and the convolutional conditional neural process (ConvCNP) \citep{Vaughan2021}. Both these methods as well as our own were fitted to the grided data detailed in Section \ref{sec:data} spanning a 20-year period, divided into a training set from 1979 to 1993 and a validation set from 1993 to 1999. We then compare their forecasting capabilities on a held-out test set of 20 years, from 1999 to 2019.\par
Remark: \emph{It is worth reiterating that our data does not include precipitation weather forecasts as a predictor, unlike the above-mentioned studies of \cite{Harris2022, Vaughan2021}. Consequently, the performance of the benchmark methods obtained in their respective articles cannot be directly compared with the performances presented in this study (of the benchmarks as well as our approach). The inclusion of precipitation NWPMs predictions would improve a model's forecasting capabilities for rainfall.}

We begin by looking at the Area Under the Curve (AUC) [Appendix \ref{apdx:diagnostics_AUC}] resulting from a calibration task, as shown in Figure~\ref{fig:compare_benchmark_a}, with corresponding Receiver Operating Characteristic (ROCs) curves reported in Appendix \ref{apdx:roc}. The AUC achieved by the Cens-JGNM is slightly higher than that of the ConvCNP, and both remain constant across different precipitation amounts. The VAE-GAN's AUC is generally lower than the two other methods and deteriorates for higher precipitation amounts, {showing that the two aforementioned models are better suited for modelling high rainfall events.} \par
 In Figure~\ref{fig:compare_benchmark_b}, we show rank histograms [Appendix \ref{apdx:diagnostics_rank}] to investigate calibration. The Cens-JGNM and ConvCNP have almost identical performances, displaying a close fit to the ideal rank, which we attribute to their similar form of marginal densities with the adoption of a Gamma mixture model. The slight misalignment of ranks on the far right for the two aforementioned approaches suggests that the parametric form is not entirely accurate in the tail, and could benefit from a more appropriate treatment of extreme values. On the other hand, the VAE-GAN displays a poor performance characteristic of under-dispersion with ranks over-represented at 0 and 1, arguing in favour of a parametric approach for marginal densities. \par
The Empirical Survival Functions (ESFs) [Appendix \ref{apdx:diagnostics_ecdf}] of each approach are studied in Figure~\ref{fig:compare_benchmark_c}. The Cens-JGNM has the closest fit to the empirical frequency, suggesting it has the best calibration out of the three models. As this diagnostic is aggregated over space, we attribute ConvCNP's overestimation of realised rainfall to the lack of explicit dependence modelling leading to miss-calibrated joint forecasts. The under-dispersion and non-explicit modelling of zero values of the VAE-GAN is exemplified with an incorrect intercept at $0 \,mm$ and a lower frequency for higher amounts.\par

\begin{figure}
\centering
        \begin{subfigure}{0.32\textwidth}
         \includegraphics[width= \textwidth]{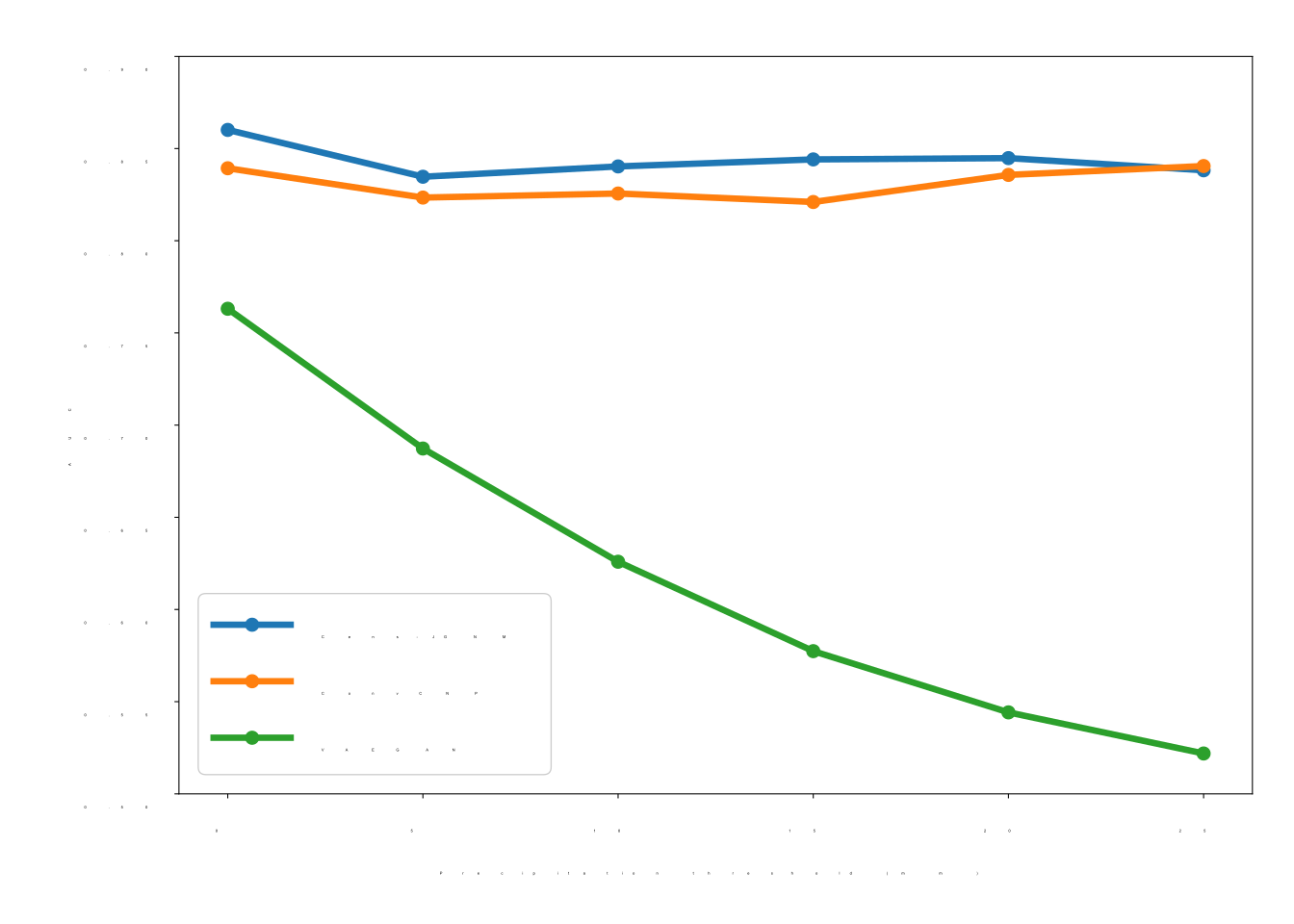}
         \caption{AUC}
         \label{fig:compare_benchmark_a}
        \end{subfigure}
    \begin{subfigure}{0.32\textwidth}
         \includegraphics[width= \textwidth]{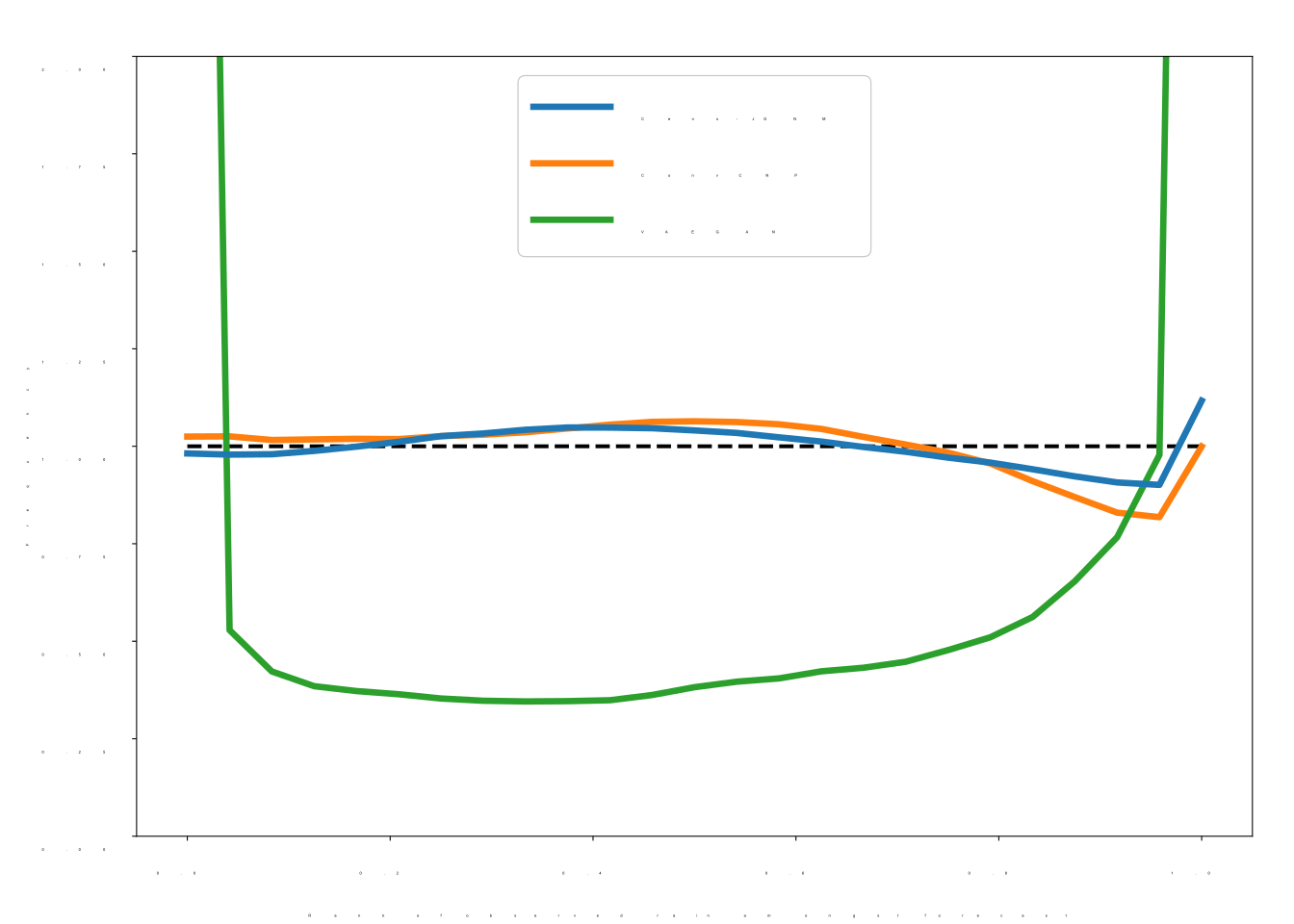}
         \caption{Rank histogram}
         \label{fig:compare_benchmark_b}
         \end{subfigure}
    \begin{subfigure}{0.32\textwidth}
         \includegraphics[width= \textwidth]{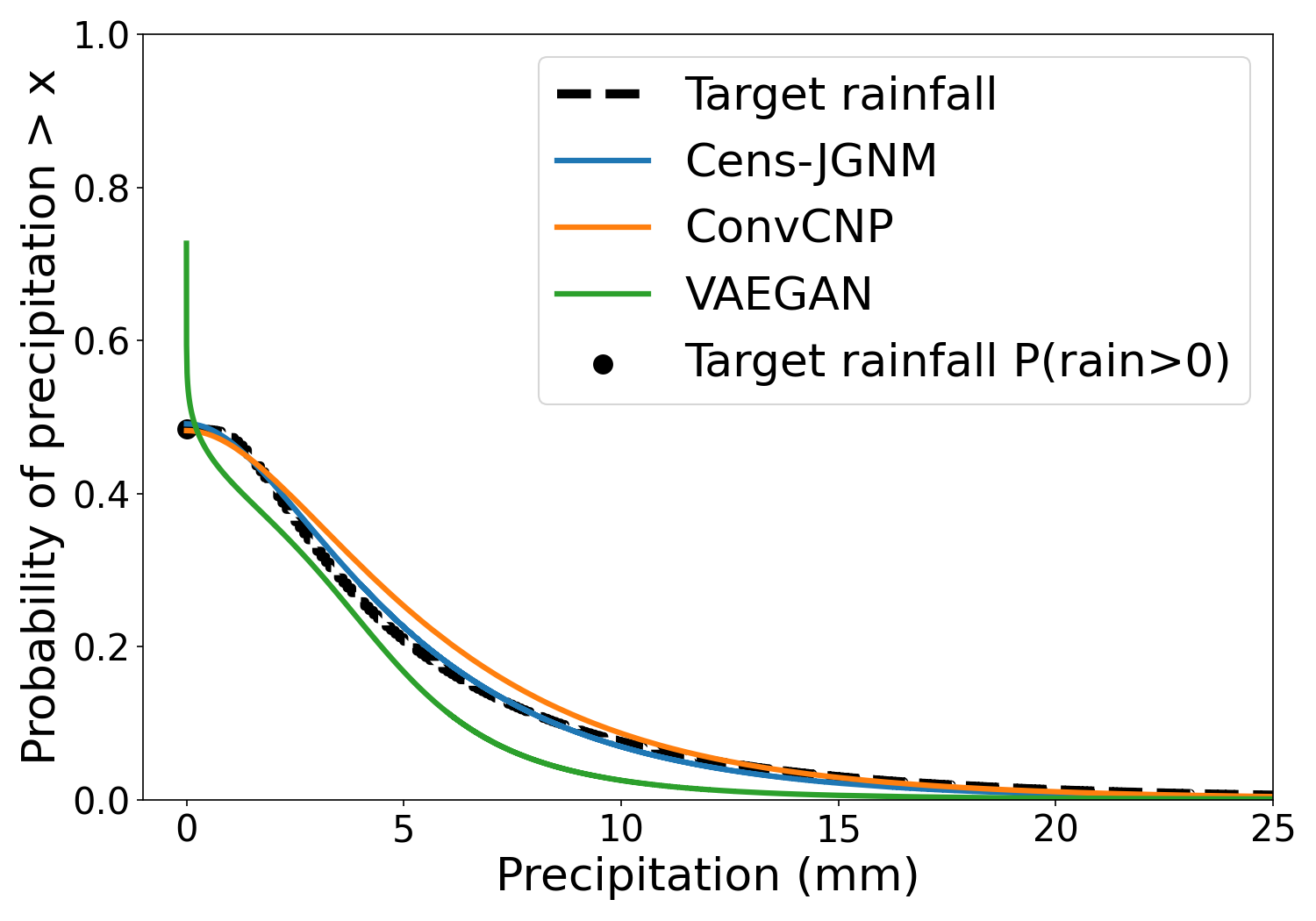}
    \caption{ESF}
    \label{fig:compare_benchmark_c}
    \end{subfigure}
    \caption{\textbf{Calibration diagnostics:} We show in (a) AUCs [Appendix \ref{apdx:diagnostics_AUC}], (b) Rank histograms [Appendix \ref{apdx:diagnostics_rank}] of the rank of precipitation observations against simulated samples and (c) ESFs [Appendix \ref{apdx:diagnostics_ecdf}] for our Cens-JGNM approach compared to ConvCNP and VAE-GAN benchmarks. Figures \ref{fig:compare_benchmark_a} and \ref{fig:compare_benchmark_b} indicate the appropriate treatment of marginal densities by the Cens-JGNM and ConvCNP both employing explicit Gamma densities as opposed to the implicit approach of the VAE-GAN. Figure \ref{fig:compare_benchmark_c} reveals that the Cens-JGNM obtains the best calibration, closely followed by the ConvCNP, with the VAE-GAN incorrectly capturing 0 values and underpredicting higher amounts.}

\end{figure}

\subsection{Spatial coherence}
\label{sec:spat_coh}

\begin{figure}
    \centering
    \includegraphics[width=0.95\linewidth]{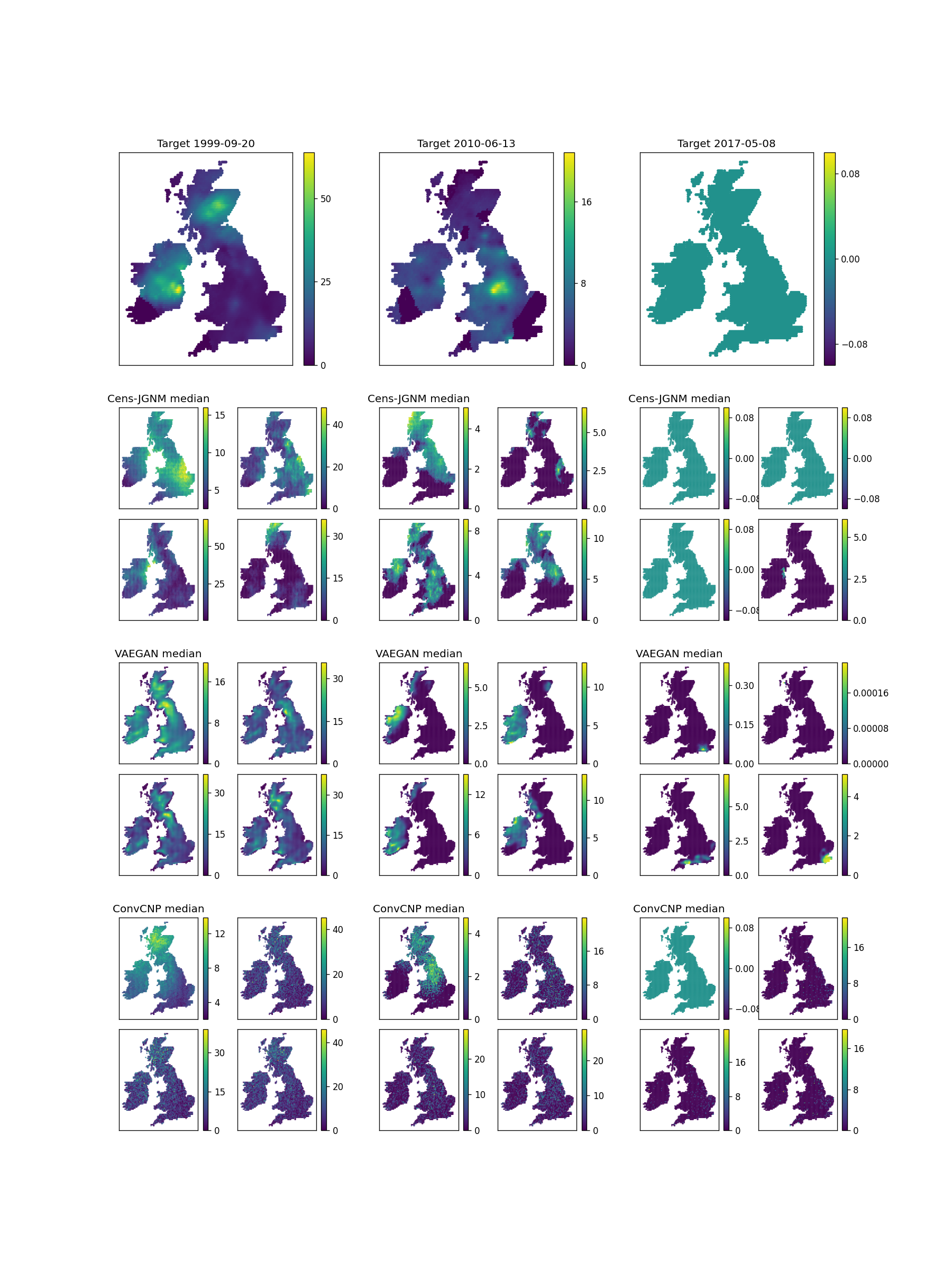}
    \caption{\textbf{Rainfall forecasts:} For a given target day at the top of each column, we show row by row samples from the Cens-JGNM, the VAE-GAN and the ConvCNP models. Each model's quadrant top left image is the median forecast, with the remaining three figures being random samples for that day. The Cens-JGNM and VAEGAN samples look spatially consistent while the ConvCNP samples are disparate across space.}
    \label{fig:samples_all}
\end{figure}

To assess the spatial fidelity of each model, we compute multiple metrics detailed in Appendix \ref{apdx:diagnostics}. Qualitatively, we also show samples from each model for three different days in Figure \ref{fig:samples_all}. Visually, the ConvCNP displays poor spatial dependence as rainfall values vary sharply between neighbouring locations. The VAEGAN and Cens-JGNM samples display adequate spatial smoothness. Notably, all models have a large variability in their samples, rarely matching the target rain exactly, possibly indicating the stochastic nature of the rainfall downscaling task.\par

\begin{figure}[t]
\centering
        \begin{subfigure}{1\textwidth}
         \includegraphics[width= \textwidth]{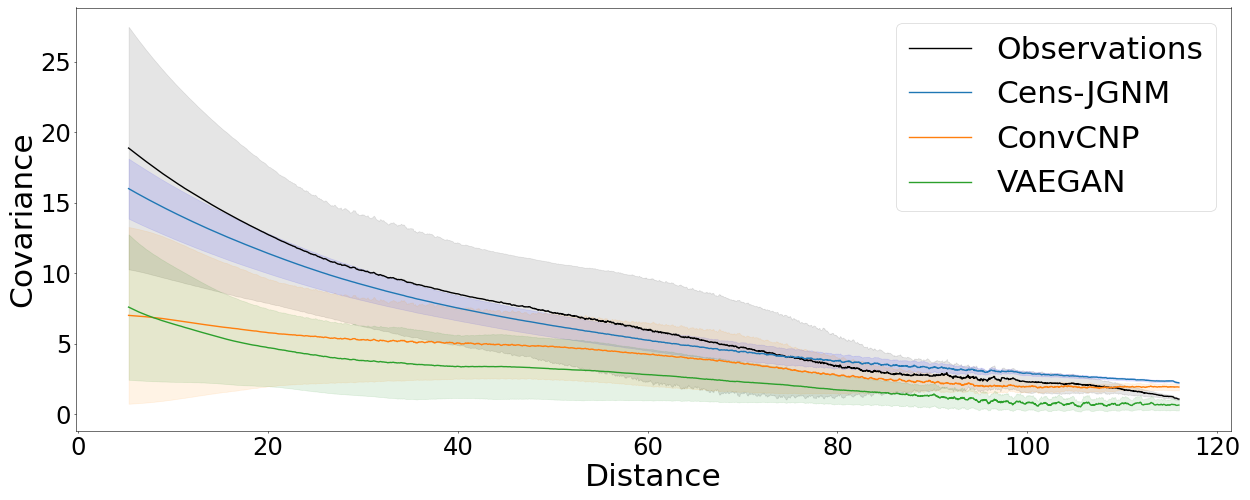}
    \caption{Covariance of rainfall against distance}
    \label{fig:benchmark_crosscorr_a}
    \end{subfigure}
    
    \begin{subfigure}{0.32\textwidth}
         \includegraphics[width= \textwidth]{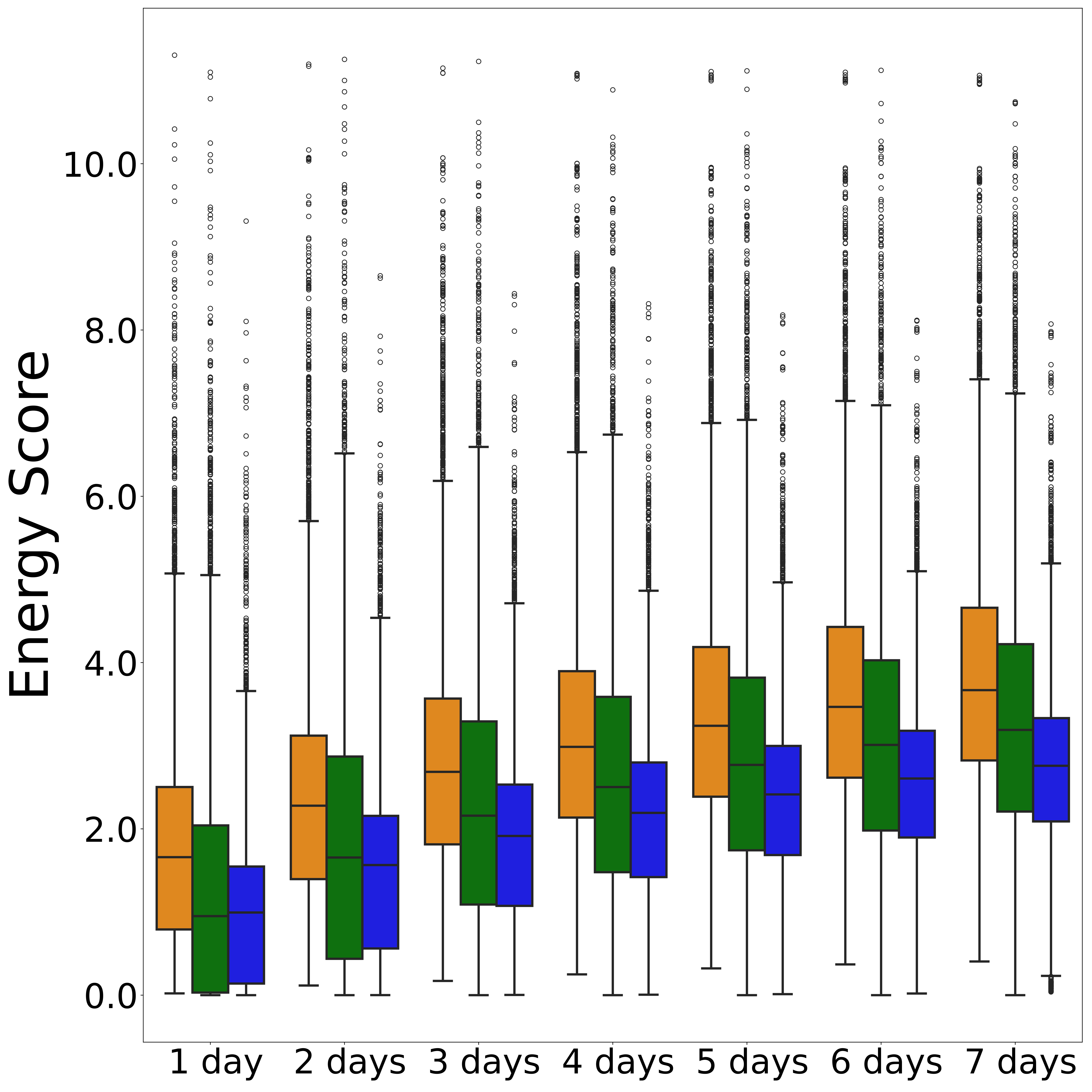}
    \caption{London}
    \label{fig:benchmark_crosscorr_b}
    \end{subfigure}
        \begin{subfigure}{0.32\textwidth}
         \includegraphics[width= \textwidth]{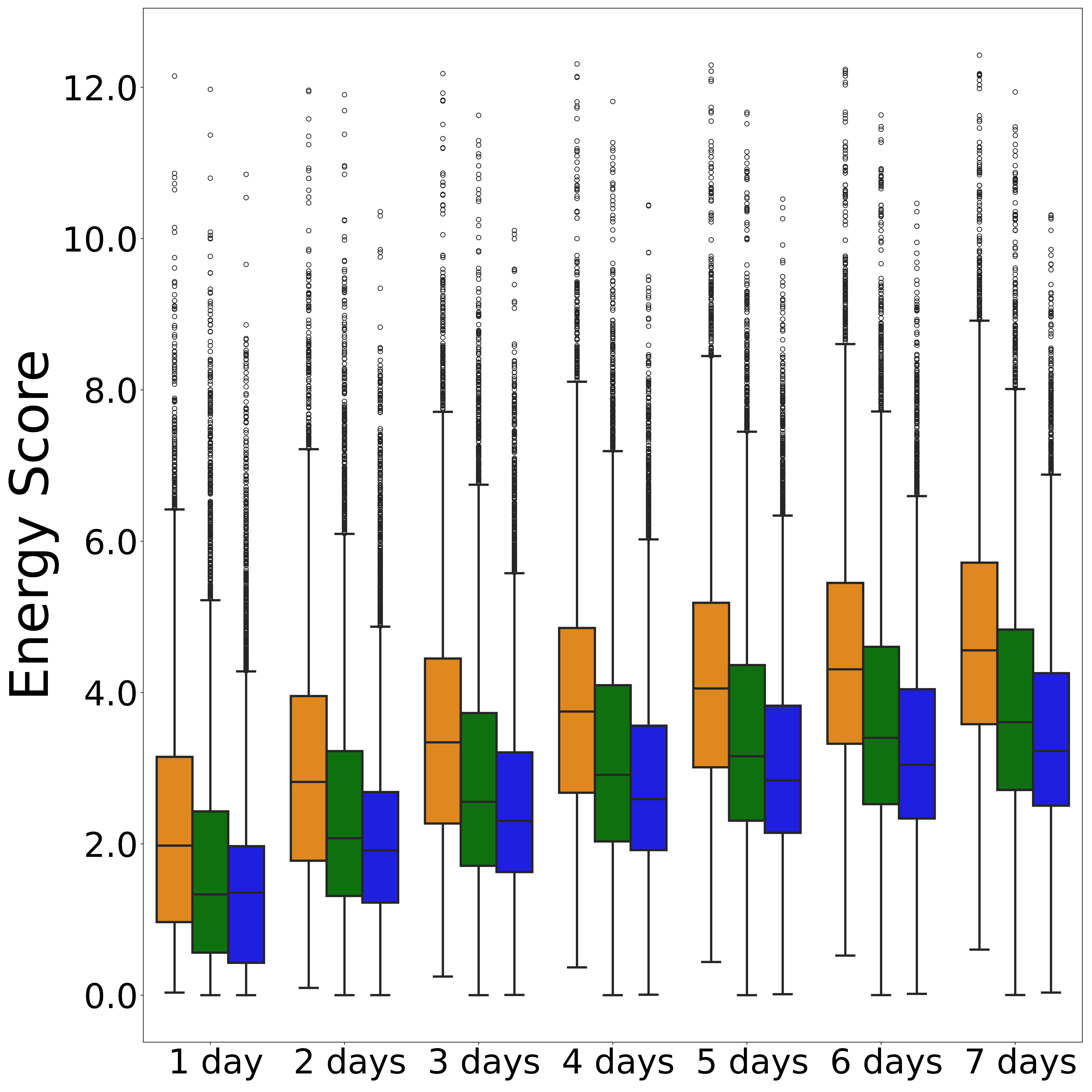}
    \caption{Glasgow}
    \label{fig:benchmark_crosscorr_c}
    \end{subfigure}
        \begin{subfigure}{0.32\textwidth}
         \includegraphics[width= \textwidth]{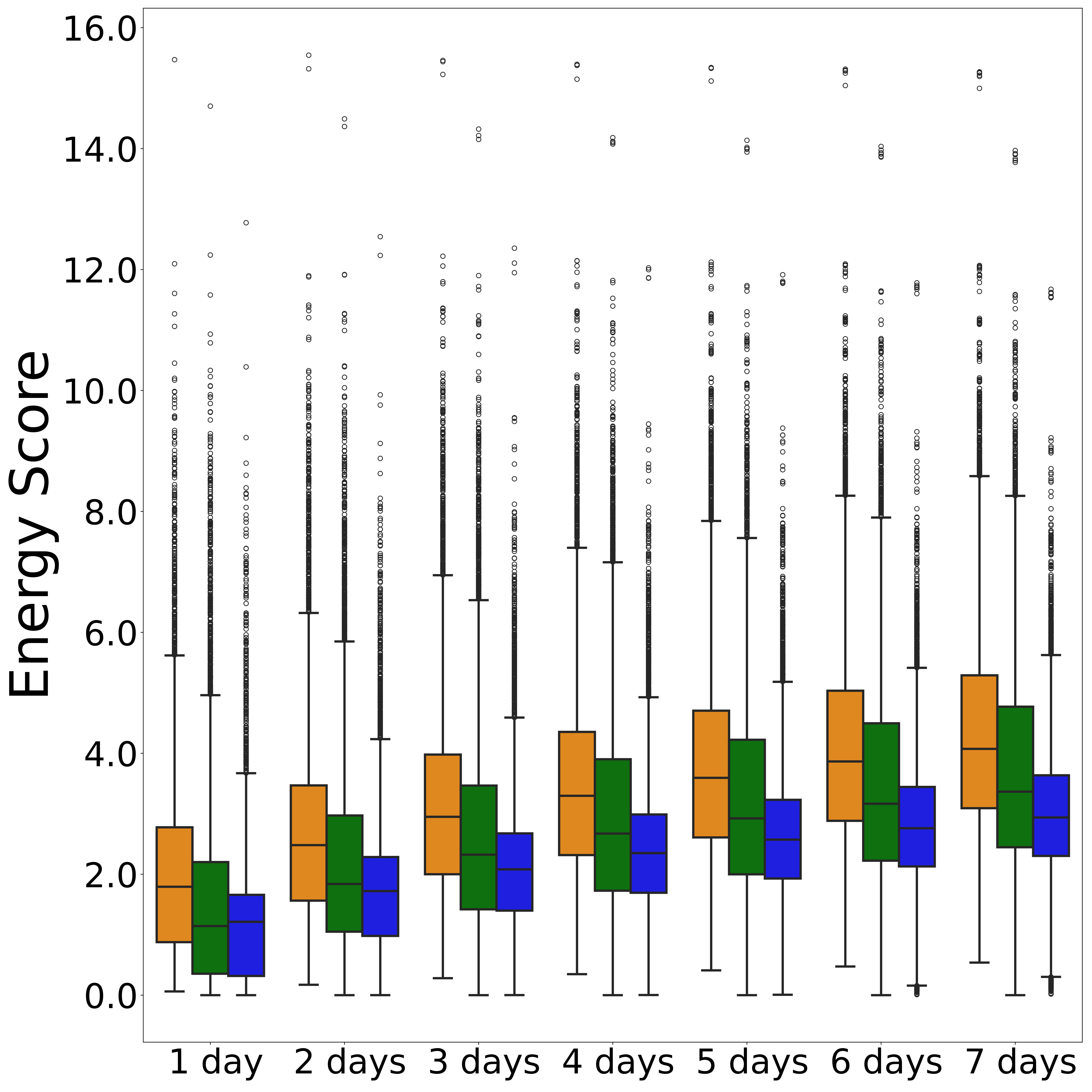}
    \caption{Dublin}
    \label{fig:benchmark_crosscorr_d}
    \end{subfigure}
    \caption{\textbf{Spatial and temporal coherence:} We show covariance of rainfall against distance (a), for observed data and all three models. We show a spatio-temporal energy score across regions of London (b), Glasgow (c) and Dublin (d). As evidenced by plot (a), our Cens-JGNM is the most effective approach at capturing spatial dependence while the VAE-GAN and ConvCNP capture minimal dependence only. Plots (b,c,d) show the stronger spatio-temporal predictive performance of our method over benchmarks.}
    \label{fig:comapre_model_spatil_crosscorr}
\end{figure}

For quantitative results, we begin by computing the average covariance across distances in simulated samples from each model and observed data, shown in Figure \ref{fig:benchmark_crosscorr_a}.{ For a given distance (on the x-axis), for all the pairs of locations $i,j$ exactly that far apart, we compute the sample covariance (on the y-axis) between the values at those two locations over all days in $\mathcal{T}$. We show the average covariance over all such pairs $i,j$ as a line with confidence bands as the standard deviation over all these pairs. Note that for higher distances, the number of locations that distance apart greatly decreases, thus leading to more variable covariance estimates despite the confidence bands shrinking.\par

The Cens-JGNM overall matches the observed covariance the closest of all models, being aligned to the observed covariance for all distances. The Cens-JGNM and ConvCNP both slightly overestimate the covariance for large distances, while the VAEGAN underestimates the covariance significantly across all distances. The standard deviation of the Cens-JGNM is lower than for observed data, meaning that our model does not greatly differentiate between different pairs and only cares about the distance between them. This is expected since distance is the only input into our copula covariance matrix which is the main factor behind the covariance.}\\

\begin{figure}[t]
\centering
         \includegraphics[width= 0.8\textwidth]{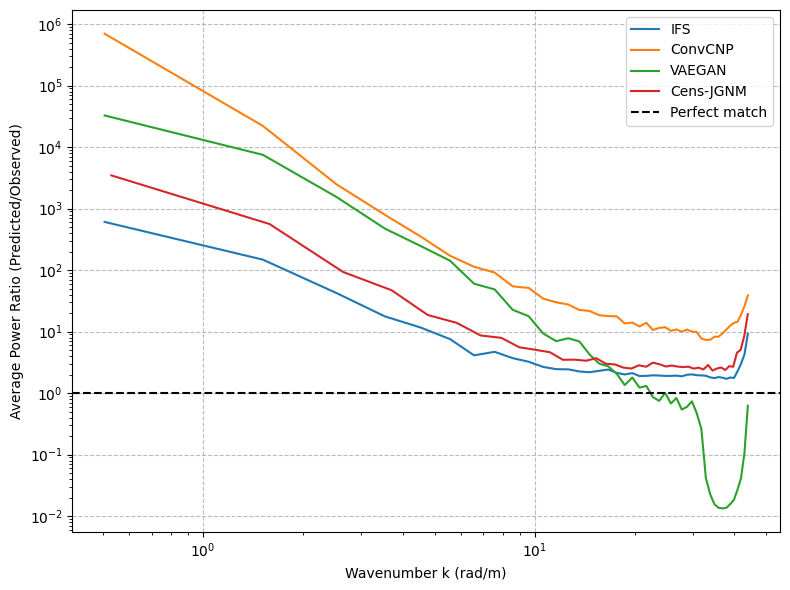}
    \caption{\textbf{Spectral Distribution:} Average ratio of spectral energies across wavenumbers be
tween predictions for each model and the true observed rainfall values. The high resolution deterministic benchmark IFS demonstrates the best overall performance, while Cens-JGNM shows superior results among neural network-based approaches and is similar to the IFS, particularly at higher wavenumbers.}
    \label{fig:spectral_plot}
\end{figure}

{Next, in Figures \ref{fig:benchmark_crosscorr_b},\ref{fig:benchmark_crosscorr_c}, and \ref{fig:benchmark_crosscorr_d}, we compute the aggregate rainfall across the set of locations surrounding London, Glasgow and Dublin respectively, over multiple days as explained in Appendix \ref{apdx:diagnostics_spatemp}. We compare the daily values to each model's forecasts using the Energy Score since these constitute multivariate data. We compute the score for all sliding windows of a given period, shown as boxplots. Aggregating the rainfall loses some spatial information, mostly retaining the specific characteristics of the region. On the other hand, using the Energy captures the temporal aspect.}\par
Across all days and locations, the Cens-JGNM performs best, followed by the VAEGAN and ConvCNP. The spread of scores is lower for the Cens-JGNM while those of the two other models are larger.

Further, for each model and time step, we compute the power spectrum of the rainfall field by applying a two-dimensional Fourier transform, squaring its magnitude, and azimuthally averaging over circular bins to obtain \(P(k,t)\) at each wavenumber \(k\). The ratio \(P_{\mathrm{pred}}(k,t)/P_{\mathrm{obs}}(k,t)\) is then calculated for each time step and subsequently averaged over the entire 20-year prediction period to yield the final curves shown in Figure~\ref{fig:spectral_plot}. In these plots, the x-axis represents the wavenumber in radians per meter (inversely related to the spatial scale), while the y-axis denotes the time-averaged power ratio, \(\bar{R}(k)\). The horizontal line at \(y = 1\) signifies an ideal match between predicted and observed spectra. Notably, the IFS \citep{ERA5}, achieves the best agreement with the observed spectra across all wavenumbers. Among the neural network-based models, our proposed method Cens-JGNM demonstrates superior performance across the spectra, with an especially strong performance at high wavenumbers 
 indicating a strong ability to effectively resolve fine-scale precipitation patterns—while both ConvCNP and VAEGAN show significant overestimation at lower wavenumbers (corresponding to larger spatial scales). \\

\begin{figure}[t]
    \centering
    \includegraphics[width=1\linewidth]{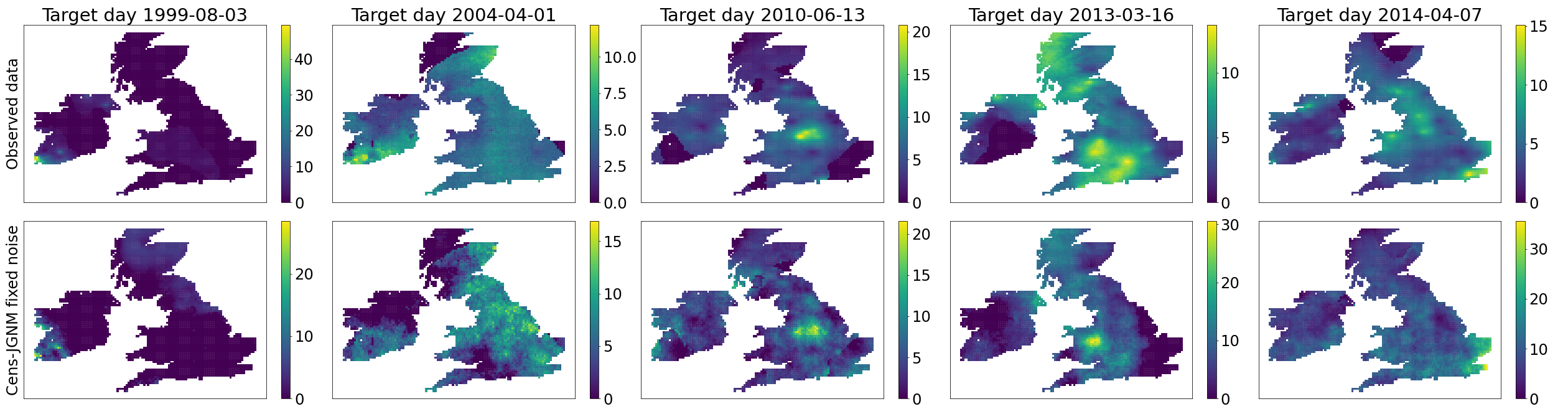}
    \caption{\textbf{Cens-JGNM samples with fixed noise inputs:} We revert observed data to decorrelated Gaussian noise using an inverse Cholesky operation. We then use this noise to sample from our Cens-JGNM (bottom row) and compare these samples to observed data (top row) across 5 different days (each column). The likeness of samples to real data provides evidence of the good fit of our covariance matrix. It also highlights the stochasticity intrinsic to rainfall downscaling.}
    \label{fig:noise}
\end{figure}

To test the correctness of our correlation matrix $\Sigma(\mathcal{D}|\theta)$, we generate samples from the Cens-JGNM using noise obtained from observed data. The idea is to use a Cholesky decomposition on the empirical covariance of observed data on the Gaussian scale to recover a decorrelated Gaussian version of the observed data. If the covariance is correctly specified, then the samples generated with the fixed observation-based noise as input will closely match the observed data. We detail this in Appendix \ref{apdx:diagnostics_noise}. We show the resulting samples in Figure \ref{fig:noise} alongside observed data for five days in the test set.\par
For each of the days, the fixed noise samples from the Cens-JGNM resemble the observed data, with higher values of rain occurring in the same regions, and other patterns roughly being matched in each column. This suggests that the correlation matrix $\Sigma(\mathcal{D}|\theta)$ is well fitted to the data.\par 
It more importantly exemplifies the inherent stochasticity of rainfall data as studied in this paper. As shown, in Figure \ref{fig:samples_all}, the samples obtained by our model are varied and do not necessarily match the observed rain. However, marginal diagnostics suggest that our marginal models are well-calibrated, meaning they should correctly capture the distribution of observed rain. Now, with Figure \ref{fig:noise}, we have shown that the copula is also well estimated, since using the ``correct" (obtained from observed data) standard Gaussian noise leads to samples that match observed data. Therefore, this suggests that observed rain is only a specific random realisation of standard noise, meaning the realised value could have been very different by pure chance; the observed data is only a glimpse of the randomness that could have been observed. This highlights the inherent stochasticity of this modelling problem, making probabilistic approaches that can capture this stochasticity essential.\\

We conclude our comparison with Table \ref{tab:benchmark}, presenting various diagnostics metrics assessing calibration, accuracy and spatial dependence. these metrics include the Continuous Ranked Probability Score (CRPS) [Appendix \ref{apdx:diagnostics_crps}], energy score [Appendix \ref{apdx:energy_score}], variogram score [Appendix \ref{apdx:diagnostics_vario}] as well as the Root Mean Square Error (RMSE) and Mean Absolute Bias (MAB) [Appendix \ref{apdx:diagnostics_rmsb_mab}]. Our Cens-JGNM model outperforms the competing probabilistic downscaling models in all the diagnostics measures, with the exception of the CRPS where the ConvCNP is slightly better. The ConvCNP focuses all modelling efforts on its marginal models which we believe is the reason for its strong performance on the CRPS. For the RMSE and MAB, our Cens-JGNM stands a close second to the IFS benchmark. This is expected, as the IFS is a challenging benchmark to outperform here as this prediction is of higher resolution than the input variables used in our study.

\vspace{1cm}
\begin{table}[H]

\centering
\begin{tabular}{|p{3cm}|p{3cm}|p{3cm}|p{3cm}|p{1cm}|}
 \hline
 \multicolumn{5}{|c|}{Diagnostic Metrics for Probabilistic Forecasting} \\
 \hline
 Model & Cens-JGNM & ConvCNP & VAE-GAN & IFS\\
 \hline
 CRPS   & 3.5631& \textbf{3.2678} &   4.1146 & --\\
 Energy Score&   \textbf{15.7253}  & 17.0146   &18.5991 & --\\
 Variogram Score & \textbf{6,862,187} & 10,333,690 &  9,091,811 & --\\
 RMSE & 3.8194 & 3.9314 &  4.4903 & \textbf{3.756}\\
 MAB & 2.1438  & 2.2179 & 2.5820 & \textbf{2.118}\\
 \hline
\end{tabular}
\caption{\textbf{Comparison with benchmarks:} Numerical diagnostics for our Cens-JGNM model against the ConvCNP, VAE-GAN models and the deterministic IFS. As IFS is a deterministic prediction, we can not compute probabilistic metrics for it. The Cens-JGNM generally outperforms other probabilistic models. For the RMSE and MAB, our model stands a close second to the IFS benchmark, which is a difficult benchmark to outperform due to its higher resolution than the inputs in our training data.}
\label{tab:benchmark}
\end{table}

\subsection{Robustness on Length of Training Data Set}
\label{sec:len_traindat}
In addition to comparing our approach against benchmarks, we appraise the robustness of the Cens-JGNM to reduced amounts of data. We train a Cens-JGNM on increasingly smaller data samples and compare the resulting forecasts on common test data. There are four Cens-JGNM versions in total, respectively fitted on 4 years of data (01/1985 to 06/1989), 6 years (1983 to 06/1989), 8 years (1981 to 06/1989) and finally 10 years (1979 to 06/1989). All are tested on the same 30 years of data, from 07/1989 to 07/2019. We use the same diagnostics as in Sections \ref{sec:comp_benchmark} and \ref{sec:spat_coh} above, to determine whether the performance of our method suffers in settings with a lack of data.\par

We begin by inspecting the AUCs (with ROCs reported in Appendix \ref{apdx:roc}) in Figure \ref{fig:robustness_a}. The different sample sizes have no significant effect, as all four versions achieve similar AUCs. The models are better calibrated for lower rainfall events, which is explained by the reduced fitting data, implying that fewer extreme observations are used to train each model.
We also compare their rank histograms in Figure \ref{fig:robustness_b}, which again show no significant difference or trend across model versions.
From Figure \ref{fig:robustness_c}, showing ESFs, it is again indicated that periods above 4 years of data are sufficient to ensure the strong performance of our model as all four versions display nearly identical performance. Importantly, the performance of these versions matches the performance of the Cens-JGNM trained on 20 years of data in Section \ref{sec:comp_benchmark}. Finally, in Table~\ref{tab:robustness_talble}, we compare the metrics of the four model versions.\par\par

\begin{figure}
\centering
        \begin{subfigure}{0.32\textwidth}
         \includegraphics[width= \textwidth]{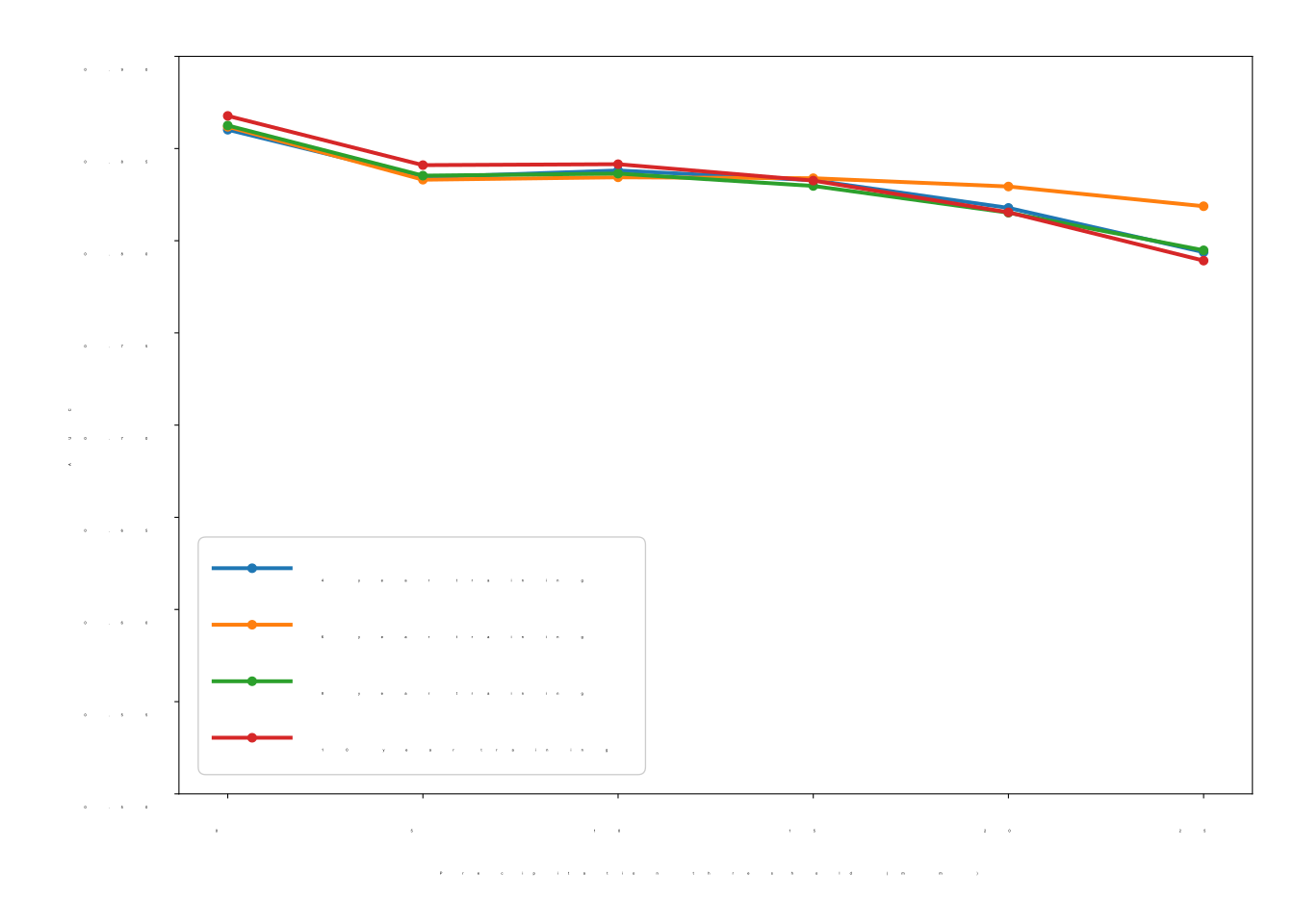}
         \caption{AUC}
         \label{fig:robustness_a}
        \end{subfigure}
    \begin{subfigure}{0.32\textwidth}
         \includegraphics[width= \textwidth]{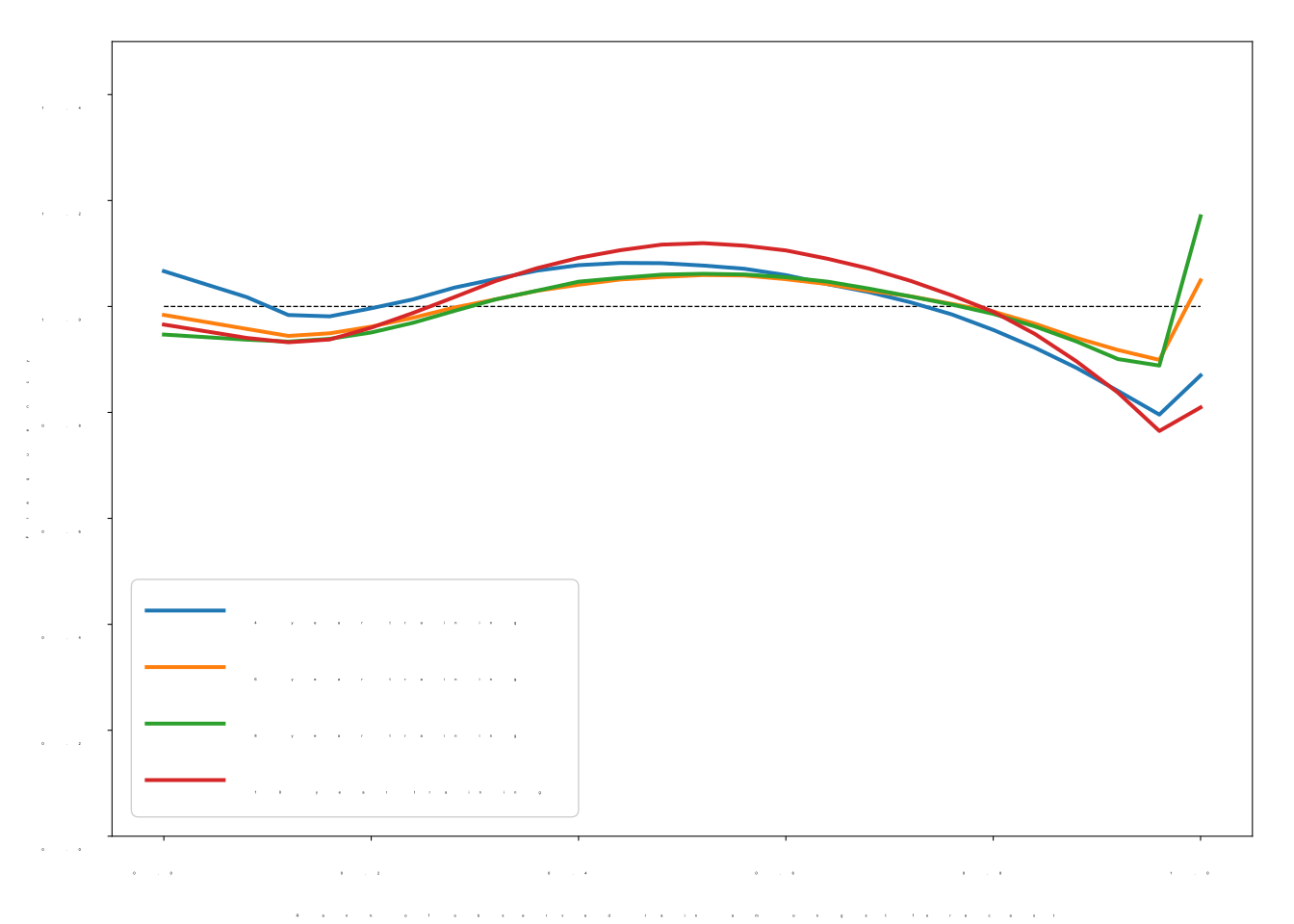}
         \caption{Rank histogram}
         \label{fig:robustness_b}
         \end{subfigure}
    \begin{subfigure}{0.32\textwidth}
         \includegraphics[width= \textwidth]{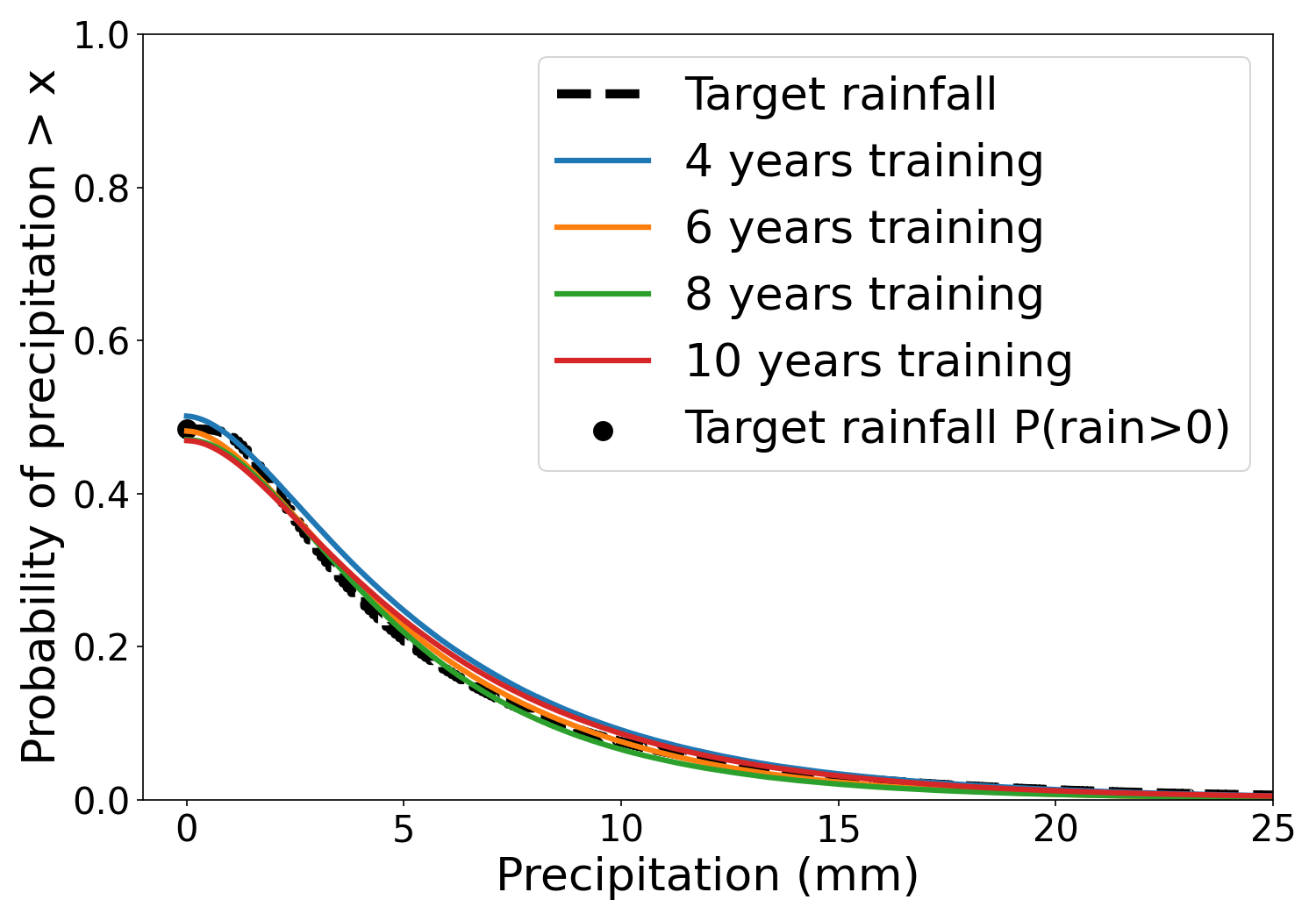}
    \caption{ESF}
    \label{fig:robustness_c}
    \end{subfigure}
\caption{\textbf{Robustness:} We show the (a) AUCs, (b) rank histograms, and (c) ESFs for the Cens-JGNM model versions fitted on decreasing amounts of data (10, 8, 6, and 4 years). The diagnostics show no trend across data size, indicating that the Cens-JGNM is robust to reduced amounts of data.}
\end{figure}

\begin{table}[H]

\centering
\begin{tabular}{|p{3cm}|p{2.25cm}|p{2.25cm}|p{2.25cm}|p{2cm}|}
 \hline
 \multicolumn{5}{|c|}{Diagnostic Metrics for Robustness Study} \\
 \hline
 Metric & 4 years & 6 years & 8 years & 10 years\\
 \hline
 CRPS   & 3.6129    & {3.4322} &   3.6996 & 3.5638\\
 Energy Score&   16.0627  & {16.0571}   & 16.0603 & {15.9226}\\
 Variogram Score & {4,918,176}  & {4,690,858} &  5,069,616 & 5,102,977\\
 RMSE    & 4.0362 & 4.0237 &  4.0275 & {3.9724}\\
 MAB&   2.2336  & 2.2336 & {2.2295} & 2.1968\\
 \hline
\end{tabular}
\caption{\textbf{Robustness:} Numerical diagnostics for the Cens-JGNM approach fitted on multiple reduced amounts of training data. The results indicate no trend in the performance of our approach across training data quantity.}
\label{tab:robustness_talble}
\end{table}


\section{Related Work}
\label{sec:lit_rev}
In this section, we review some existing literature connected to our work. We begin in Section \ref{sec:lit_rev_marginal} by discussing other methods of forecasting rainfall in a univariate context. Next, in Section \ref{sec:lit_rev_copula}, we explore the estimation issues present in copulas when the likelihood becomes complicated to evaluate or directly unavailable. Finally, we describe alternative approaches to probabilistic rainfall forecasting than our two-step procedure in Section \ref{sec:lit_rev_others}.\par

\subsection{Marginal rainfall estimation}
\label{sec:lit_rev_marginal}

In \cite{paper10}, a study on GLMs used for rainfall density estimation with Markovian assumptions on temporal dependence recommends the use of a Bernoulli-Gamma mixture or an ensemble model, motivating our JGNM parametric choice. This comparison study further reinforces the findings of \citep{das1955fitting,fealy2007statistical,ailliot2015stochastic,holsclaw2017bayesian,bertolacci2019climate,xie2023assessment}, all reporting success with marginal models based on Gamma densities. PCA regression was used in \cite{paper11} for precipitation downscaling using CM variables together with mean sea level pressure (MSLP) measurements, alleviating the underprediction problems faced by CMs. Our JGNM can be seen as a generalisation, where the refinement of variables used in a regression model is made in a data-driven way. Quantile regression is used by \cite{paper12} to get daily precipitation distributions by conditioning on CM outputs, increasing the flexibility in the selection of predictors, which our JGNM increases even further. \cite{su2019evaluation} applied Bayesian model averaging (BMA) and stepwise regression models (SRMs) to downscaling. Our approach incorporates a pre-treatment of predictors as in SRMs with uncertainty quantification as in BMA, combining both strengths under a single approach.\par 
Deep Learning (DL) methods have been used effectively in Geosciences, such as \cite{li2021deep} applying DL methods to forecast location-wise densities of solar energy. In \cite{2021Trunet}, expanding on previous DL models applied to rainfall forecasting such as \cite{ConvLSTMshi,DeepSD,miao2019improving} and \cite{pan2019improving}, the authors achieved state-of-the-art performance in point-wise prediction, showing that DL methods can be used to downscale rainfall based on CM forecasts. We directly improve on their approach by incorporating uncertainty quantification in the marginals.\par

\subsection{Copula estimation problems}
\label{sec:lit_rev_copula}
As is common when the copula is latent or in the presence of zero-inflated data (both being the case in our work), MLE-based inference becomes complicated or unavailable and alternatives need to be considered.\par
For extreme rainfall modelling, \cite{huser2019modeling} assume a censored copula for extreme values, targeting the full likelihood for inference, but only for up to 15 locations, noting that a high dimension is a limiting factor for Gaussian copulas, reinforcing the conclusions of \cite{huser2017bridging}. Modelling extreme rainfall through a Brown–Resnick process, \cite{richards2021spatial} resort to a composite likelihood approach as an alternative to a computationally unfeasible MLE. More recently, in \cite{richards2022modelling} and \cite{richards2023joint}, authors use a censored Gaussian copula model for the dependence of extreme rainfall events, circumventing the unavailable likelihood by adopting a pseudo-likelihood approach with spatially-informed sub-sampling. Outside of rainfall forecasting, \cite{dobra2011copula} model graphical binary and ordinal variables in a Bayesian framework with a latent Gaussian copula, requiring an approximation to the likelihood function with the extended rank likelihood of \cite{d2007extending}. Thus, as was summarised in \cite{huser2016likelihood}, while the use of these pseudo-likelihoods allows for inference, performing inference in higher dimensions remains an open problem. Our minimum scoring estimate approach provides a direct solution by not relying on likelihoods, additionally ensuring exact and computationally efficient parameter inference.\par

In the presence of zero-inflated data, \cite{yoon2020sparse} introduce a truncated latent Gaussian copula as an analogue to censored copulas and develop an estimator for the latent correlation matrix expanding on a rank-based procedure from \cite{fan2017high}. This approach is expanded by \cite{chung2022phylogenetically} to a Bayesian setting on graphical models for zero-inflated count-data data where inference is performed using Gibbs sampling. While their approach allows for higher dimensions, it is framed inside the Bayesian paradigm, whereas our minimum scoring rule approach is exact and can be used in both frequentist and Bayesian settings (with extensions following the works of \cite{pacchiardi2021generalized}).\par
Recent advances in likelihood-free inference have led to the development of alternative methods for copulas inference. \cite{alquier2022estimation} use the maximum mean discrepancy (MMD) as a function to minimise in order to learn the parameters of copulas, obtaining a covariance estimator robust to miss-specification. Similarly, \cite{janke2021implicit} employ the energy distance (a possible instance of the MMD) as a loss function in the training of a generative network to approximate a copula distribution, avoiding the necessity for hyperparameters for the MMD. Our work extends their approaches to a spatial setting while accounting for censored observations.\par

\subsection{Alternative probabilistic forecasting approaches for rainfall}
\label{sec:lit_rev_others}

Non-homogeneous hidden Markov models for rainfall occurrence are considered by \cite{paper9} with extensions for amounts by \cite{charles1999spatiotemporal}, modelling occurrence and amount of rainfall separately, which we do jointly with our JGNM. We further avoid their conditional independence assumptions on the spatial dependence structure with our censored copula approach. \cite{frost2011comparison}, a review of six downscaling methods (namely \cite{chandler2002glimclim,timbal2004southwest,mcgregor2005c,mehrotra2007preserving,chiew2009estimating} and a HMMs-based \citep{charles1999spatiotemporal} implementation of \cite{kirshner2007learning}), found that HMMs displayed weaknesses in capturing spatial correlation. Applying Bayesian HMMs to rainfall, \cite{song2014bayesian} (expanding on the approach of \cite{paper11}) uses a covariance matrix construction with spatial distances used only in error correlation. \cite{holsclaw2017bayesian} assume a more complicated time-dependency in transition probabilities, which calls for a Polya-Gamma-based MCMC sampling scheme. \cite{bertolacci2019climate} use a spatial structure captured through a Gaussian process with marginal densities modelled by mixtures of expert models. Our approach extends these methods with direct and richer spatial modelling through the censored latent Gaussian copula.\par

Ensemble Model Output Statistics (EMOS), introduced by \cite{gneiting2005calibrated} and improved by the works of \cite{dabernig2017spatial,moller2016probabilistic,moller2020probabilistic}, models rainfall using an appropriate density of which parameters are estimated with linear regression equations incorporating spatio-temporal effects. \cite{rasp2018neural} use neural networks instead of linear regression equations, further improved by \cite{mlakar2023ensemble}, performing estimation jointly across all future times and locations. Our approach similarly estimates distributions jointly in a spatio-temporal manner, however, we additionally employ the censored latent copula to impose the correct spatial dependence for any given forecast. \par
Machine learning approaches include \cite{watson2020investigating}, with the use of Generative Adversarial Networks (GANs) to increase the resolution of rainfall forecasts while simultaneously aiming to correct biases. This work was expanded upon by \cite{price2022increasing} by using a conditional GAN based on coarse weather variables. Similarly, \cite{Harris2022} substitutes the GAN generator with a variational auto-encoder for conditional forecasts, adapted to cyclone-induced rainfall by \cite{vosper2023deep}. Alternatively, \cite{pacchiardi2022probabilistic,chen2022generative} use generative neural networks with the energy score as a loss function to avoid the problematic training of GANs. All these approaches learn an implicit likelihood and suffer from drawbacks such as the inability to generate zero values. Our JGNM model ensures explainability and a proper fitting of zero values and we compared our spatial dependence to the study of \cite{Harris2022}, demonstrating improved performance. Further, our model can be evaluated on new locations after training, unlike most of these methods. \cite{Vaughan2021} employ a convolutional conditional neural process model with parametric marginals, explicit modelling of zero values and inclusion of out-of-sample locations. Their approach is very similar to the JGNM, although their model fails at capturing spatial dependence to the same extent as our censored copula method, as evidenced in Section \ref{sec:comp_benchmark}. \cite{ascenso2023improving} use a convolutional network approach with a modified loss function to enforce the spatial coherence of rainfall. We have a similar focus on spatial dependence with the energy score but additionally provide uncertainty quantification and interpretability.


\section{Conclusion} 
\label{sec:ccl}

In this paper, we introduced a novel method for spatio-temporal 
downscaling of zero-inflated data. We described our first component of the model, JGNM as a distributional learner capturing temporal trends in conjunction with surrounding information to describe distributions for quantities of interest at a location. The JGNM is very flexible and can be evaluated at any location, even one outside of the learning set. We also introduced censored latent Gaussian copulas as an answer to the spatial dependence modelling of censored data. We explained the intractability of the likelihood of this model, prohibiting the use of regular MLE-based methods. Hence we offered a scoring-rule-based inferential scheme, resulting in cheaper and more robust parameter inference. Our censored latent Gaussian copula methodology allows for the automatic inclusion of new locations, making our complete approach capable of accommodating additional locations added to the model without requiring parameter re-estimation. We compared our model to competitive benchmarks focusing on the same task \citep{Vaughan2021,Harris2022}. We demonstrated our model's ability to correctly capture uncertainty to the same or better extent as competitors. Furthermore, we showcased our model's superior capture of spatial dependence over benchmarks, rendering our approach better suited for scenarios where spatial coherence is desired.\par
This work is an important improvement towards more reliable post-processing techniques as well as precipitation downscaling capable of correctly capturing uncertainty and spatial dependence. Moreover, while the main motivation of our work is rainfall downscaling, our methodology can effortlessly be applied to any spatio-temporal probabilistic prediction, with or without censoring. Therefore, we believe that our work is not only of interest to the weather science community but indeed relevant to the broader multivariate forecasting community. Further, we have used model fields as predictor variables, which can be resolved reasonably well by typical climate models, hence one possible future direction could be to use the developed methodology for post-processing of climate models.\par
Another possible extension of our model is the construction of the correlation matrix of the latent censored Gaussian copula in Section \ref{sec:cens_copula_spadep}. Specifically, in the initial distance $\mathcal{D}$, one could include information on differences in weather variables to better capture the spatial dependence. In fact, it is possible to extend our copula approach to explicitly capture temporal dependence by adapting the kernel parameter according to temporal information. Other improvements particular to the modelling of extreme rainfall events could be made to the JGNM by considering extreme value distributions \citep{behrens2004bayesian,macdonald2011flexible,ding2019modeling,gao2021spatial,richards2022modelling}, either entirely or by adding them as a mixture term in the tail. Finally, we acknowledge that imposing a Gaussian correlation structure, while computationally appealing and leading to easy spatial modelling, might be too restrictive, especially for extreme variations in rainfall levels at nearby locations. In this context, we could consider other types of copulas, no longer restricted to Gaussian dependence structures \citep{janke2021implicit,wadsworth2022higher,richards2023joint}.


\section*{Acknowledgements and  Disclosure of Funding}
D.H. is funded by the Center for Doctoral Training in  Mathematical Sciences at Warwick. R.A is funded by the University of Warwick and Southern University of Science and Technology. R.D. is funded by EPSRC (grant nos. EP/V025899/1 and EP/T017112/1) and NERC (grant no. NE/T00973X/1).

\bibliographystyle{abbrvnat}

\newpage
	\section*{Appendix}
\appendix

\section{Network Architecture}
\label{apdx:JGNM_NN}

Our joint generalized neural model (JGNM) is designed to predict rainfall by parameterizing a three-parameter distribution. The model takes as input 6-hourly weather fields from the ERA5 reanalysis and predicts three parameters (\(p\), \(\mu\) and \(\phi\) ) for a predictive distributions for daily precipitation at a given point on a high-resolution grid.

Specifically, the input data consists of low-resolution weather fields 
\[
\boldsymbol{\mathcal{X}}_j \in \mathbb{R}^{20 \times 21 \times 6}, \quad j = 1,\ldots,112,
\]
where each \(\boldsymbol{\mathcal{X}}_j\) represents one 6-hour time step over a 28-day period (4 time steps per day). These fields span a \(20 \times 21\) grid covering the region of interest and include 6 weather variables (specific humidity, air temperature, geopotential height at 500 hPa, eastward and northward wind components at 850 hPa, and total column water vapour). Note that precipitation is deliberately excluded from the input.

Following the approach in \cite{2021Trunet}, we first perform bilinear interpolation to map the input from the coarse \(20 \times 21\) grid to a finer \(100 \times 140\) grid, matching the 0.1° resolution (approximately 8.5 km) of the observed rainfall data. Due to computational limits, we then partition the interpolated grid into \(16 \times 16\) sub-sections which form the input to our neural network. For each sub-section, the 6-hourly data for all 6 variables is organized into a sequence of matrices 
\[
\boldsymbol{x}_j \in \mathbb{R}^{16 \times 16 \times 6}, \quad j = 1,\ldots,112.
\]
Thus, the input for a 28-day period is represented as a sequence of shape \((112, 16, 16, 6)\).

The model first applies a Time Distributed 2D Convolutional Layer (TD2L) that processes each \(\boldsymbol{x}_j\) independently with the same 2D convolution, transforming it into a hidden representation 
\[
\boldsymbol{h}^{\text{TD2L}}_j \in \mathbb{R}^{16 \times 16 \times 64}.
\]

Next, the sequence of hidden representations is fed into a stacked Convolutional LSTM (ConvLSTM) module. This module consists of two bi-directional ConvLSTM layers with skip connections, which capture both spatial and temporal dependencies in the data. In our experiments, the ConvLSTM layers are configured with an input dropout of 0.25 and a recurrent dropout of 0.35.

Following the ConvLSTM, a Time Distributed Temporal Downscaling (TDTD) layer reduces the sequence length from 112 (6-hourly intervals) to 28 (daily values) by applying the same downscaling operation on non-overlapping groups of 4 time steps. The TDTD layer integrates a convolutional self-attention (CSA) module, a Gaussian Error Linear Unit (GeLU) activation \citep{hendrycks2016gaussian}, and two 2D convolutional layers to effectively aggregate temporal information.

Finally, the output from the TDTD is passed through a Time-Distributed Mean Function Layer. This layer comprises three single-layer perceptrons, each acting as an inverse link function mapping the extracted features to one of the three distribution parameters. Consequently, the model generates parameter predictions:

\[
\mu_i, \phi_i, p_i \in \mathbb{R}^{16 \times 16}, \quad i = 1, \dots, 28,
\]

which represent daily predictive distributions of total rainfall at each location within the \(16 \times 16\) grid, corresponding to the initial inputs \( \boldsymbol{x}_j, j = 1, \dots, 112 \). Finally, these spatially segmented \(16 \times 16\) sub-images are combined to form a complete spatial prediction over the entire target region.

A schematic of the overall architecture is shown in Figure~\ref{fig:NeuralNetStructure}. 
\begin{figure}[ht]
    \centering
    \includegraphics[width=0.4\textwidth]{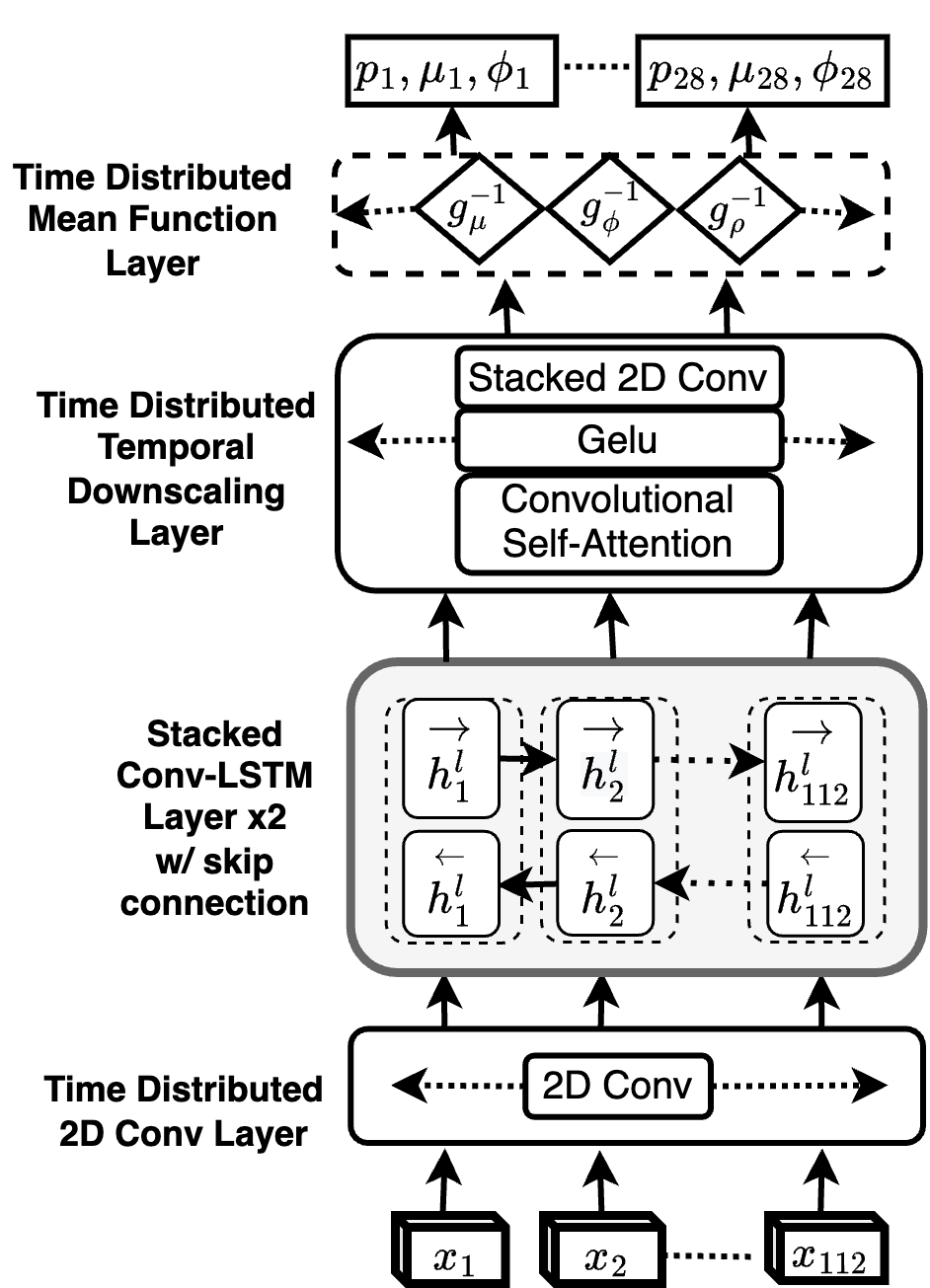}
    
    \caption{\textbf{Joint Generalised Neural Model (JGNM) architecture.} The JGNM takes as input a sequence of \(112\) tensors of shape \(16\times16\times6\), corresponding to 6-hourly weather fields over 28 days. After initial processing by a Time-Distributed 2D Convolutional Layer (TD2L), the data passes through stacked bidirectional ConvLSTM layers capturing spatiotemporal dependencies. A Time-Distributed Temporal Downscaling Layer (TDTD) then aggregates temporal information from 6-hourly intervals to daily intervals, reducing sequence length from \(112\) to \(28\). Finally, a Time-Distributed Mean Function Layer outputs daily predictive parameters \(p_i, \mu_i, \phi_i \in \mathbb{R}^{16\times16}\), parameterizing a predictive zero-gamma mixture distribution for total rainfall at each grid point. These \(16\times16\) predictions are subsequently combined into the full spatial grid of size \(100\times140\).}

    \label{fig:NeuralNetStructure}
\end{figure}

\subsection{Convolutional Self-Attention (CSA)}
\label{apdx:ConvMHSA}
The Convolutional Self-Attention (CSA) module is designed to reduce the temporal resolution of a sequence of three-dimensional feature maps, preserving maximal information from the original input while simultaneously incorporating long-range dependencies. In our implementation, the CSA module transforms an input tensor
\[
x \in \mathbb{R}^{(b, t_{\text{input}}, h, w, c)},
\]
where \(b\) is the batch size, \(t_{\text{input}}\) is the initial sequence length, \(h\) and \(w\) denote the spatial dimensions, and \(c\) is the number of channels, into an output tensor with a reduced temporal dimension \(t_{\text{output}} = t_{\text{input}} / s\) (with \(s\) the downscaling factor).

\paragraph{Positional Embedding and Sub-sequence Formation:}  
The input tensor is first partitioned into non-overlapping sub-sequences of length \(s\). For each sub-sequence 
\[
x_{[j:j+s]} \in \mathbb{R}^{(b, s, h, w, c)},
\]
we flatten the spatial dimensions to obtain
\[
\tilde{x}_{[j:j+s]} \in \mathbb{R}^{(b, s, d)} \quad \text{with } d = h \times w \times c.
\]
A learnable temporal positional embedding \(P \in \mathbb{R}^{(1, s, d)}\) is then added to each sub-sequence to incorporate information about the order of the elements:
\[
x'_{[j:j+s]} = \tilde{x}_{[j:j+s]} + P.
\]

\paragraph{Query, Key, and Value Generation:}  
For each sub-sequence \(x'_{[j:j+s]}\), we compute the attention components as follows. The query vector is obtained by averaging the sub-sequence along the temporal dimension:
\[
q_{[j:j+s]} = \frac{1}{s} \sum_{t=j}^{j+s-1} x'_t \in \mathbb{R}^{(b, d)}.
\]
The key and value tensors are taken as the reshaped (but unaveraged) sub-sequence:
\[
k_{[j:j+s]} = \tilde{x}_{[j:j+s]} \in \mathbb{R}^{(b, s, d)}, \quad
v_{[j:j+s]} = \tilde{x}_{[j:j+s]} \in \mathbb{R}^{(b, s, d)}.
\]

\paragraph{Multi-Head Projection:}  
To enable the model to capture multiple types of relationships, we perform multi-head attention. For each head \(h \in \{1, \ldots, H\}\), the query, key, and value are linearly projected:
\[
Q^h = q_{[j:j+s]} W^Q_h, \quad K^h = k_{[j:j+s]} W^K_h, \quad V^h = v_{[j:j+s]} W^V_h,
\]
where \(W^Q_h,\, W^K_h,\, W^V_h \in \mathbb{R}^{d \times d_{\text{hidden}}}\) are learned projection matrices. This results in:
\[
Q^h \in \mathbb{R}^{(b, d_{\text{hidden}})}, \quad K^h,\, V^h \in \mathbb{R}^{(b, s, d_{\text{hidden}})}.
\]

\paragraph{Scaled Dot-Product Attention:}  
For each head, attention weights are computed using scaled dot-product attention:
\[
\text{Attention}^h = \text{softmax}\!\left(\frac{Q^h (K^h)^\top}{\sqrt{d_{\text{hidden}}}}\right) V^h,
\]
where the softmax is applied over the \(s\) (temporal) dimension, yielding an output of shape \(\mathbb{R}^{(b, d_{\text{hidden}})}\) for each head.

\paragraph{Aggregation and Output Projection:}  
The outputs from all \(H\) heads are concatenated:
\[
\text{Concat}\big(\text{head}_1, \ldots, \text{head}_H\big) \in \mathbb{R}^{(b, H \cdot d_{\text{hidden}})},
\]
and then projected with a learned matrix \(W^O \in \mathbb{R}^{(H \cdot d_{\text{hidden}}, d_{\text{output}})}\) to produce the final sub-sequence representation:
\[
x_{[j:j+s]}^{\text{(output)}} = \text{Concat}\big(\text{head}_1, \ldots, \text{head}_H\big) W^O.
\]

\paragraph{Reshaping to Recover Spatial Dimensions:}  
Finally, the output for each sub-sequence is reshaped back to the original spatial dimensions, resulting in a tensor
\[
x^{\text{(output)}} \in \mathbb{R}^{(b, t_{\text{output}}, h, w, c)},
\]
which represents the temporally downscaled sequence of feature maps.

In our implementation, we set \(h = 4\), \(w = 4\), \(c = 6\), \(s = 4\), \(d_{\text{hidden}} = 64\), \(d_{\text{output}} = 64\), and use \(H = 4\) attention heads. As $t_{\text{input}}$

\subsection{ Time Distributed Mean Function Layer }
\label{apdx:GLMMeanFunction}

In Table~\ref{tab:mean_function_gamma}, we list the inverse link functions (mean functions) used in the Time Distributed Mean Function Layer for the Zero-Gamma mixture density. Note that these functions serve as activation functions mapping the real-valued output of the neural network back to the target parameter space. For example, with reference to our link functions introduced in Section~\ref{sec:JGNM}, the inverse of \(g_{\mu}\) for \(\mu\) is implemented as
\[
g_{\mu}^{-1}(x) = \left(6.0\cdot\text{ReLU}(x+0.4)\right)+1\times10^{-6},
\]
which ensures \(\mu\in[0,\infty)\) while maintaining numerical stability. A similar formulation holds for \(\phi\) via the inverse of \(g_{\phi}\). 
\begin{table}[!htbp]
\begin{center}
\begin{tabular}{lll}
\hline
       & Parameter Range & $g^{-1}$ : Mean (Inverse Link) Function                           \\ \hline
\(p\)      & \((0,1)\)         & \(\text{sigmoid}(x)\)   \\
\(\mu\)  & \([0,\infty)\)    & \(\left(6.0\cdot\text{ReLU}(x+0.4)\right)+1\times10^{-6}\)\\
\(\phi\) & \([0,\infty)\)    & \(\left(6.0\cdot\text{ReLU}(x+0.5)\right)+1\times10^{-6}\)
\end{tabular}
\caption{\textbf{GLM Mean Functions:} Inverse link (activation) functions used in the Time Distributed Mean Function Layer of the Zero-Inflated JGNM.}
\label{tab:mean_function_gamma}
\end{center}
\end{table}

\section{Construction of Distances Matrix}
\label{app:dist_const}

Within a latent Gaussian copula, one needs to estimate a correlation matrix $\Sigma$, which we construct as a function of an initial distance matrix $\mathcal{D}$. As we want the copula to be spatially informed, we consider the latitude and longitude of locations as well as the topography given by geopotential height at a given location. For a given parameter value $a \in \mathbb{R}_{\geq0}$, we construct the distance matrix $\mathcal{D}^a$ by computing for all pairs of locations $i,j$:
\begin{equation*}
    \mathcal{D}_{(i,j)}^a= \sqrt{|lat_i-lat_j|^2+|lon_i-lon_j|^2+|a \cdot (topo_i-topo_j)|^2}.
\end{equation*}
Here, $a$ serves to adequately weigh the importance of topography on the resulting correlation matrix $\Sigma=k(\mathcal{D}^a|\theta)$. With $a\to0$, our copula model relies on geographical distances only while $a\to\infty$ makes the dependence rely uniquely on topography.\par

In a preliminary experiment, we experimented with optimising $a,\theta$ jointly, as well as setting $a$ at pre-defined values (chosen at $a=0,0.005$, and $0.03$) to only optimise $\theta$. We show the Energy score values for different setups in Table \ref{tab:energy_distmat} computed using forecasts after each optimisation set-up.

\begin{table}[ht]
    \centering
    \begin{tabular}{c|cccc}
        \toprule
        Set-up & $a=0.0$ &$a=0.005$ & $a=0.03$ & $a=0.07226$ (joint optimisation) \\
        \midrule
        Energy Score &15.7253 & 15.7394 & 15.7382 & 15.7852 \\
        \bottomrule
    \end{tabular}
    \caption{Energy score values for different values of $a$}
    \label{tab:energy_distmat}
\end{table}

The best Energy score was obtained from the set-up where the topography was ignored by setting $a=0$. The joint optimisation was unstable and obtained the worst score among all setups. We believe the reason for this issue was the presence of collinearity between topography and geographical distances, as locations close by will also share similar elevations. We show samples for two days (each row) from each optimisation outcome (each column) and target rain for the day, below in Figure \ref{fig:samples_distmat}. The samples were obtained with fixed noise coming from a Cholesky decomposition on observed rain, see Appendix \ref{apdx:diagnostics_noise}.\par 
The realism flaws of relying excessively on topography become obvious. For higher values of $a$, the regions with rainfall are no longer determined by geographic closeness but rather by sharing the same elevations. This is particularly identifiable for the areas of Scotland which as a mountainous region ends up having the model misrepresent the dependence of samples.

\begin{figure}[t]
    \centering
    \includegraphics[width=\linewidth]{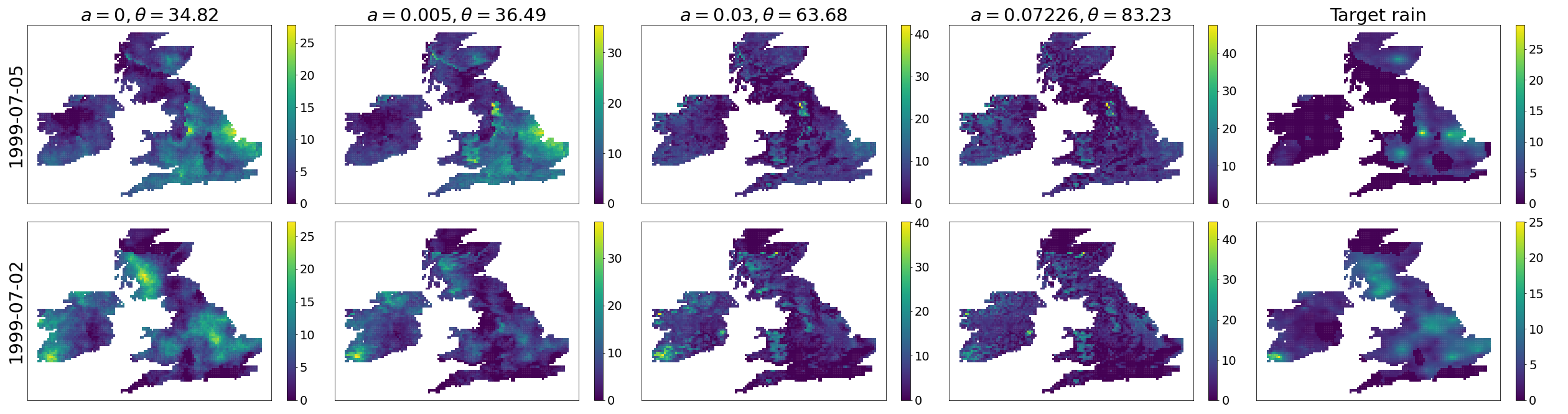}
    \caption{\textbf{Samples with varying levels of topographical dependence:} We optimise $\theta$ for various fixed values of $a$ as well as a joint optimisation set-up, and show samples for two days row-wise. The samples with higher topographical dependence are no longer realistic and over-rely on topography.}
    \label{fig:samples_distmat}
\end{figure}

\section{Scoring Rules as a Divergence} 
\label{app:SR_div}
Assume we have observed data from distribution $\mathcal{P}^{*}$ and want to choose a parameter $\boldsymbol\theta$ which parameterises a second distribution $\mathcal{P}^{\boldsymbol\theta}$ such that the two distributions are as close to each other as possible. A divergence $D$ is then defined as a function of two distributions such that (i) $D(\mathcal{P}^{*}||\mathcal{P}^{\boldsymbol\theta})\geq0$ and (ii) $D(\mathcal{P}^{*}||\mathcal{P}^{\boldsymbol\theta})=0 \iff \mathcal{P}^{*}=\mathcal{P}^{\boldsymbol\theta}$. Therefore, a divergence can be used to optimise a parameter to recover the best possible model $\mathcal{P}^{\boldsymbol\theta}$ for the data generating distribution $\mathcal{P}^{*}$ .\\

One possible choice of divergences is scoring rules (SRs). As defined in \cite{doi:10.1198/016214506000001437}, a scoring rule $S(\mathcal{P},\mathbf{z})$ is a function between a distribution $\mathcal{P}^{\boldsymbol \theta}$ and observed data $\mathbf{z}$ as a realisation of a random variable $\mathbf{Z}\sim \mathcal{P}^{*}$. Then the \emph{expected scoring rule} is defined as $S(\mathcal{P},\mathbf{z}):=\mathbb{E}_{\mathbf{Z}\sim\mathcal{P}^{*}} S(\mathcal{P}^{\boldsymbol \theta},\mathbf{Z}) $. The SR is termed \emph{proper} if relative to a set of distributions $\mathbf{P}$, if the expected SR is minimised when $\mathcal{P}^{*}=\mathcal{P}^{\boldsymbol\theta}$:
\begin{equation*}
    S\left(\mathcal{P}^{*}, \mathcal{P}^{*}\right) \leq S\left(\mathcal{P}^{\boldsymbol \theta}, \mathcal{P}^{*}\right) \,\, \forall \,\mathcal{P}^{\boldsymbol \theta}, \mathcal{P}^{*} \in \mathbf{P}. 
\end{equation*} 
Furthermore, a SR is termed \emph{strictly proper}, if the minimisation above is unique:
\begin{equation*}
    S\left(\mathcal{P}^{*}, \mathcal{P}^{*}\right) < S\left(\mathcal{P}^{\boldsymbol \theta}, \mathcal{P}^{*}\right) \,\, \forall \,\mathcal{P}^{\boldsymbol \theta}, \mathcal{P}^{*} \in \mathbf{P} \text{ s.t. } \mathcal{P}^{*} \neq \mathcal{P}^{\boldsymbol\theta}.
\end{equation*}
By considering the quantity $D_{SR}(\mathcal{P}^{*}||\mathcal{P}^{\boldsymbol\theta}):= S\left(\mathcal{P}^{*}, \mathcal{P}^{\boldsymbol \theta}\right)-S\left(\mathcal{P}^{*}, \mathcal{P}^{*}\right) $ for a strictly proper SR, one can see that it defines a divergence. Indeed, (i) is verified by the SR being proper and (ii) is verified by the additional requirement of being strictly proper. This permits the use of strictly proper SRs as divergences to perform inference on parameters of a distribution. For the choice of SR, we introduce the Energy Score as:
\label{apdx:energy_score}
\begin{equation*}
S_{\mathrm{E}}\left(\mathcal{P}^{\boldsymbol\theta}, \mathbf{z}\right) = 
2 \cdot \mathbb{E}_{\mathbf{Z}^{\prime} \sim \mathcal{P}^{\boldsymbol\theta}}\|\mathbf{Z}^{\prime}-\mathbf{z}\|_2^{\beta}-\mathbb{E}_{\mathbf{Z}_{1}^{\prime},\mathbf{Z}_{2}^{\prime} \sim \mathcal{P}^{\boldsymbol\theta}}\left\|\mathbf{Z}^{\prime}_{1}-\mathbf{Z}^{\prime}_{2}\right\|_2^{\beta}
\end{equation*}

where $\beta \in (0,2)$ is a hyperparameter regulating the severity of the divergence for incorrect distributions $\mathcal{P}^{\boldsymbol\theta}$. The energy SR is a strictly proper SR for the class of $\mathbf{P}$ such that $\mathbb{E}_{{\mathbf{Z}^{\prime}} \sim P}\|{\mathbf{Z}^{\prime}}\|^\beta<\infty$, see \cite{doi:10.1198/016214506000001437} and is also known as a rescaling of the energy distance mentioned in section 4.2. It is possible to obtain unbiased estimates of $S_{\mathrm{E}}\left(\mathcal{P}^{\boldsymbol\theta}, \mathbf{z}\right)$ by repeatedly sampling $\mathbf{z}^{\prime}$ $m$ times from $\mathcal{P}^{\boldsymbol\theta}$, as:
\begin{equation*}
\hat{S}_{\mathrm{E}}\left(\{\mathbf{z}_j^{\prime}:1\leq j \leq m\}, \mathbf{z}\right) =  \frac{2}{m} \sum_{j=1}^m\left\|\mathbf{z}_j^{\prime}-\mathbf{z}\right\|_2^\beta-\frac{1}{m(m-1)} \sum_{\substack{ j,k=1 \\ k \neq j}}^m\left\|\mathbf{z}_j^{\prime}-\mathbf{z}_k^{\prime}\right\|_2^\beta
\end{equation*}

where $\mathbf{z}$ are observations and $\mathbf{z}^{\prime}$ are samples. As such, we have a method for inferring parameters of a distribution by comparing the observations to simulated samples and minimising $D_{SR}$ using the unbiased estimate of the SR. This is equivalent to choosing $\boldsymbol{\theta^{*}}= \argmin\limits_{\boldsymbol\theta} \hat{S}_{\mathrm{E}}\left(\mathcal{P}^{\boldsymbol\theta}, \mathbf{z}\right) $ since the first part of $D_{SR}$ is constant in $\boldsymbol\theta$.

\section{Optimisation procedure}
\label{apdx:optim}
Here, we detail the training of $\theta$ from Section \ref{sec:cens_copula} as done in practice in experiments. \par
We choose a Matern kernel with $\nu=0.5$, as for this value its expression simplifies \cite{guttorp2006studies} into:

\begin{equation*}
    \Sigma(\mathcal{D};\theta)_{(i,j)} =\text{exp}(-\mathcal{D}_{(i,j)}/\theta)
\end{equation*}

letting us implement a custom version of the kernel in Pytorch \citep{paszke2019pytorch} which allows for efficient gradient-based optimisation of our copula model. \\

To initialise the optimisation, we choose a $\theta$ value which minimises the Frobenius norm of $\Sigma(\mathcal{D};\theta)-\hat{\Sigma}$ where $\hat{\Sigma}$ is the empirical covariance matrix of the observed rainfall transformed to the Gaussian scale $z_{i,s} = \Phi^{-1}(F_{i,s}(y_{i,s}|\boldsymbol{\mathcal{H}}_s))$. This serves as an initial guess with a somewhat similar correlation matrix to that of the censored data and led to a more stable optimisation in initial experiments. In particular, for certain combinations of $\mathcal{D}$ and $\theta$, the resulting correlation matrix had many entries close to zero, which caused vanishing gradients in the Scoring Rule loss. Choosing an initialisation in this way prevented this issue in our experiments.\\
Throughout our whole optimisation, we take a fixed standard Gaussian sample with 103 realisations over all locations of size (103, $\mathcal{I}$). We then compute the correlation matrix with a given $\theta$ from our optimisation step and perform a Cholesky decomposition on the correlation matrix. We then multiply the decomposed matrix with the fixed noise sample to obtain a sample from an $|\mathcal{I}|$-dimensional Gaussian with covariance matrix $\Sigma(\mathcal{D};\theta)$. We then compute the Scoring Rule loss over minibatches of 103 days by comparing observed rainfall on the Gaussian scale to the censored samples obtained through a Cholesky decomposition of the matrix, following Section \ref{sec:cens_copula_minSR}. We perform 500 gradient steps with the Adam optimiser \citep{kingma2014adam}, with a learning rate of 0.2 and with gradient clipping.\\

The resulting optimisation converges for all experiments with a smooth optimisation surface based on the trace of parameter estimates and Scoring rules throughout iterations. The optimisation is not challenging as the parameter $\theta$ is one-dimensional and takes about 3 hours on a single Intel Xeon Platinum 8260 (Cascade Lake) CPU. 

\section{Description of Diagnostics}
Here we describe the different diagnostics used.
\label{apdx:diagnostics}

\subsection{L1 Norm of Matrix differnces}
\label{apdx:diagnostics_L1}
To compare two matrices, we use the L$_1$ norm for matrices. This is simply the sum of absolute entry-wise differences between the matrices. For matrices $A,B$, it is expressed as:
\begin{equation*}
\Sigma_{i=1,j=1}^{|\mathcal{I}|}|A_{(i,j)}-B_{(i,j)}|
\end{equation*}

\subsection{AUC of calibration task}
\label{apdx:diagnostics_AUC}
We seek to verify the accuracy of our model's daily estimated probabilities for a given amount of rainfall compared to the observed amount. To achieve this, we assess the sensitivity of the detection of heavy precipitation. We convert our forecasting task into a classification task by asking the following question: “Will there be precipitation exceeding a given amount?” The answer to this question will depend on a probability threshold, which our model has to surpass in order to emit a signal, or rather, give a positive answer to the question. This threshold $\tau$ is subjective and can be chosen as any value in [0, 1].\\
For a given threshold, a test that correctly predicts heavy rain (as in rain exceeding a given amount) is known as a true positive. But if the test predicts heavy rain on a day it did not occur, this is known as a false positive.
By treating precipitation at each day and point in space as separate independent events, a true positive rate (the proportion of points and days with heavy rain which was correctly detected) and the false positive rate (the proportion of points and days with no heavy rain with a positive signal) can be obtained.\\
In order to detect heavy rain, we compare the threshold $\tau$ with $1 - F_{i,t}(q)$, that is 1 minus the CDF for location $i$ at time $t$ evaluated at the heavy rain value $q \in \mathbb{R}_{\geq0}$.
The higher the value of $\tau$, the more mass under the PDF on the right of $q$ we need
in order to be confident enough to issue a positive signal. On the contrary, assuming
$\tau = 0$, we then classify all forecasts irrespective of location, time, or predictors as a
positive.\\
By changing this threshold $\tau$, we can construct a receiver operating characteristic
(ROC) curve plotting the true and false positive rates at each of these thresholds.
Different levels of precipitation $q$ can be tested, for example, 5 mm for light rain up
to 25 mm for extreme events. The area under the ROC curve (AUC) can be used
to assess how well the prediction for different levels of precipitation in the face of
uncertainty was captured by the model. The closer the AUC gets to 1, the more
confidence we can have in the heavy rainfall prediction capabilities of the examined
model.\par
\subsection{Rank histogram} 
\label{apdx:diagnostics_rank}
Another way of assessing the calibration of a density is through rank histograms. This consists of computing the percentile value associated with an observation under a model, the percentiles being called the observation ranks. {For the Cens-JGNM and ConvCNP models, as we have explicit models for marginal densities for every location $i\in\mathcal{I}$, we can compute the ranks of observed rain directly using the model's CDF corresponding to location $i$ and day $t$. In the case of $0$-valued observations, we sample uniformly a value from the censored range $[0,1-p_{i,t}]$ following the explicit modelling of $0$. For the VAEGAN model, we sample $m$ times from the model to obtain an empirical distribution $\hat{F}_{i,t}$ for days $t$ and location $i$. Then, we can estimate $F_{i,t}( y_{i,t})$ as $\hat{F}_{i,t}( y_{i,t}) = \sum_{k=1}^{m} \mathds{1}\{  y_{i,t} >  y_{i,t}^{(k)} \} $, that is the percentile of the observed rainfall with respect to our model's simulations $\{ y_{i,t}^{(k)}:k \in \{1,\ldots,m\}\}$. We note that to break ties in ranks of $0$-valued rain, for the VAE-GAN, we add a meteorologically insignificant amount of rain to observations and simulations, following the original VAEGAN paper \citep{Harris2022}.} By obtaining the rank of observations across days, we can construct a histogram of ranks. If the model is perfectly calibrated, the ranks will be uniformly distributed on $[0,1]$, as should be the case for a CDF of its own realisations. We can assess our model's properties by inspecting the dissimilarities of the histogram to a flat line \citep{hamill2001interpretation}. For instance, if the model is over-dispersed, the ranks will tend to group in the middle while under-dispersion will manifest itself as most ranks falling within the extremes of the histogram. Equally, one can determine the bias of the model by observing unequal proportions of ranks on either side of the histogram, where an agglomeration on the right is indicative of under-prediction while agglomeration on the left indicates over-prediction.\par

\subsection{Empirical Survival Function (ESF)}
\label{apdx:diagnostics_ecdf}

 To further assess the calibration of our model, we test whether the estimated frequency of a type of event (eg. precipitation $> x$) over the testing period has matched the observed frequency. The estimated frequency of achieving a rainfall realisation above a level $x \in \mathbb{R}_{\geq 0}$ is obtained by generating samples from our model and looking at the proportion of samples exceeding $x$. This is then compared to the actual observation frequencies for values higher than $x$. A well-calibrated model's sample frequency should follow the observation line closely, while a poorly calibrated model will deviate significantly from it. We refer to these plots as empirical survival function plots as they correspond to an approximation of $1-F_{i,s}(y_{i,s}|\mathcal{H}_s)$. \par

\subsection{CRPS}
\label{apdx:diagnostics_crps}
We also rely on numerical metrics to evaluate the effectiveness of our approach. Firstly, we consider the continuous ranked probability score (CRPS) \citep{matheson1976scoring}, written as
\begin{equation*}
    \operatorname{CRPS}(F_{i,t}, y_{i,t})=-\int_{-\infty}^{\infty}(F(y_{i,t}^{'})-\mathbf{1}\{y_{i,t}^{'} \geq y_{i,t}\})^2 \mathrm{~d} y_{i,t}^{'}
\end{equation*}
which can be shown to be equal to
\begin{equation*}
    \operatorname{CRPS}(F_{i,t}, y_{i,t})=\frac{1}{2} E_{Y_1,Y_2\sim F_{i,t}}\left|Y_1-Y_2\right|-E_{Y\sim F_{i,t}}|Y-y_{i,t}|.
\end{equation*}
The above expectations can be estimated without bias through repeated sampling from the given model. The CRPS is a generalisation of the mean absolute error and indeed simplifies to it when considering a model with point-wise predictions. This metric is applied location-wise and favours well-calibrated forecasts.\par

\subsection{Variogram score}
\label{apdx:diagnostics_vario}
To specifically target the spatial coherence of our model, we resort to the Variogram score, written as:

\begin{equation*}
    S_{\mathrm{v}}^{(p)}\left(\mathbf{F}_{t}, \mathbf{y}_{t}\right):=\sum_{k, l=1}^n w_{kl}\left(\left|y_{k,t}-y_{l,t}\right|^p-\mathbb{E}_{\mathbf{Y} \sim \mathbf{F}_{t}}\left|Y_{k,t}-Y_{l,t}\right|^p\right)^2.
\end{equation*}
Again, we can appeal to repeated sampling of $n$ samples in order to estimate the expectation. We take $p=1$ for simplicity. The weights $w_{kl}$ are chosen to represent the spatial nature of the problem. As such, we choose $w_{k,l}$ to be 1 over the distance from location $k$ to $l$ and setting $w_{kk}=0$. This score takes into account the whole joint distribution and compares the smoothness of observed data against that of model forecasts. An ideal forecast would have a Variogram score of zero.

\subsection{RMSE and MAB}
\label{apdx:diagnostics_rmsb_mab}
Finally, we assess the models' performances by two quantitative metrics:  the root mean squared error (RMSE) and mean absolute bias (MAB). In order to compute these quantities, we need to convert our probabilistic forecasts into point-forecasts. We do this by relying on the median of $m$ forecasts for any given place and time. We can then compute the two metrics as:
$$
\mathrm{RMSE}=\sqrt{\frac{1}{n \times T} \sum_{i=1}^{n} \sum_{t=1}^{T}\left(y_{i,t}-\widehat{y}_{i,t}\right)^{2}}, \quad \mathrm{MAB}=\frac{1}{n \times T} \sum_{i=1}^{n} \sum_{t=1}^{T}\left|y_{i,t}-\widehat{y}_{i,t}\right|
$$
where $\widehat{y}_{i,t}$ is the median for time $t$ and location $i$, $y_{i,t}$ is {the recorded
amount of rainfall at time $t$ and location $i$}, and $T$ is the total number of days considered.

\subsection{Spatio-Temporal Energy Score}
\label{apdx:diagnostics_spatemp}
To assess a model's predictive performance for flood events, we test its effectiveness on aggregate rainfall prediction over a given region of locations $\mathcal{I}'\in\mathcal{I}$. Specifically, we compute the aggregate rain over all locations $i\in \mathcal{I}'$ for all days $s$ in a period of days $s+=(s,s+1,...)$, ranging from 1 to 7 days. This gives us one multivariate realization of aggregate daily rainfall for multiple days $\mathbf{y_{s+}}$, which we can compare against forecasted values using $m$ model samples $\{y_{s+,j}'\}_{j=1}^m \sim\mathcal{P_\theta}$ with the Energy Score of Appendix \ref{apdx:energy_score} (in the case of 1 day, this reduces to the CRPS \citep{gneiting2005calibrated}). We compute this energy score for all sliding windows of $s+$ days in our experiments. The expression is as follows: 
\begin{equation*}
    SR_\text{spa-temp}(\mathcal{P_\theta},\mathbf{y_{s+}}):=\hat{\mathcal{S}}_E(\{y_{s+,j}'\}_{j=1}^m,\mathbf{y_{s+}}).
\end{equation*}

\subsection{Spectral Energy Plot}
\label{appdx:spectral_en_plot}
Mathematically, we define the time-averaged spectral ratio as

\[
\bar{R}(k) = \frac{1}{T} \sum_{t=1}^{T} \left| \frac{P_{\mathrm{pred}}(k,t)}{P_{\mathrm{obs}}(k,t)} \right|,
\]

where \(T\) is the total number of time steps. Here, the power spectrum \(P(k,t)\) for a field \(f(x,y,t)\) is computed from its 2D Fourier transform:

\[
P(k,t) = \left\langle \left|\mathcal{F}\{f(x,y,t)\}\right|^2 \right\rangle_{\sqrt{k_x^2+k_y^2}=k},
\]

with the averaging \(\langle \cdot \rangle\) taken over all Fourier components corresponding to the wavenumber \(k\). This concise formulation quantifies the average proportional difference in spectral energy between the predicted and observed fields across spatial scales.

\subsection{Samples with fixed decorrelated noise.}
\label{apdx:diagnostics_noise}
Here we describe the experiment shown in Figure \ref{fig:noise} of the main text. We begin by transforming observed data to the Gaussian scale, using 
\begin{equation*}
    z_{i,s} = \Phi^{-1}(F_{i,s}(y_{i,s}|\boldsymbol{\mathcal{H}}_s)).
\end{equation*}

We then compute the empirical covariance of these $\mathbf{z}_s=(z_{1,s},\ldots,z_{n,s})$ observations, call it $\hat{\Sigma}$. Then, we compute $\tilde{\mathbf{z}}=\mathbf{z}\cdot (Chol(\hat{\Sigma}))^{-1}$ where $Chol$ represents a Cholesky decomposition, which (assuming $\hat{\Sigma}$ is the correct covariance) will give decorrelated standard Gaussian data. Next, we use this $\tilde{\mathbf{z}}$ as an input to a Cholesky decomposition-based-sampling with our $\Sigma(\mathcal{D}|\theta)$, to obtain $\tilde{\mathbf{u}}=\Phi\left(\tilde{\mathbf{z}}\cdot Chol(\Sigma(\mathcal{D}|\theta))\right)$, which (assuming $\tilde{\mathbf{z}}$ is standard Gaussian) will be a sample from the latent Gaussian copula with correlation matrix $\Sigma(\mathcal{D}|\theta)$. We then transform this back to the data scale with $\tilde{y}_{i,s}'=F_{i,s}^{-1}(u_{i,s})$. \par

Assuming that $\hat{\Sigma}$ is correct, this generated sample $\tilde{y}_{i,s}'$ should match closely the observed rainfall $y_{i,s}$ across all $i$. If $\Sigma(\mathcal{D}|\theta)$ and the covariance of the data closely match, it is straightforward to see that they should cancel in the transformations used above, meaning the generated samples should be identical. As such, the closer the samples, the more evidence there is that the correlation matrix of the copula model is correctly fit.

\section{ROC for Experiments}
\label{apdx:roc}

\begin{figure}[ht]
\centering
        \begin{subfigure}{0.3\textwidth}
         \includegraphics[width= \textwidth]{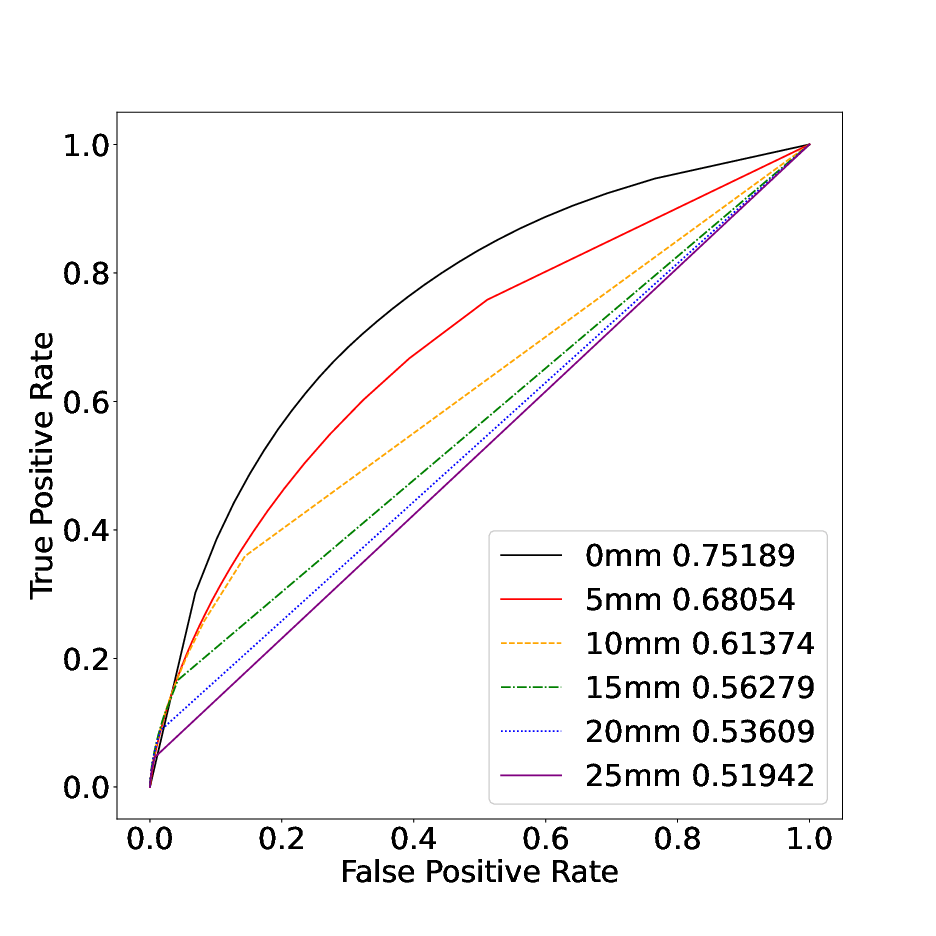}
    \caption{VAE-GAN}
    \end{subfigure}
        \begin{subfigure}{0.3\textwidth}
         \includegraphics[width= \textwidth]{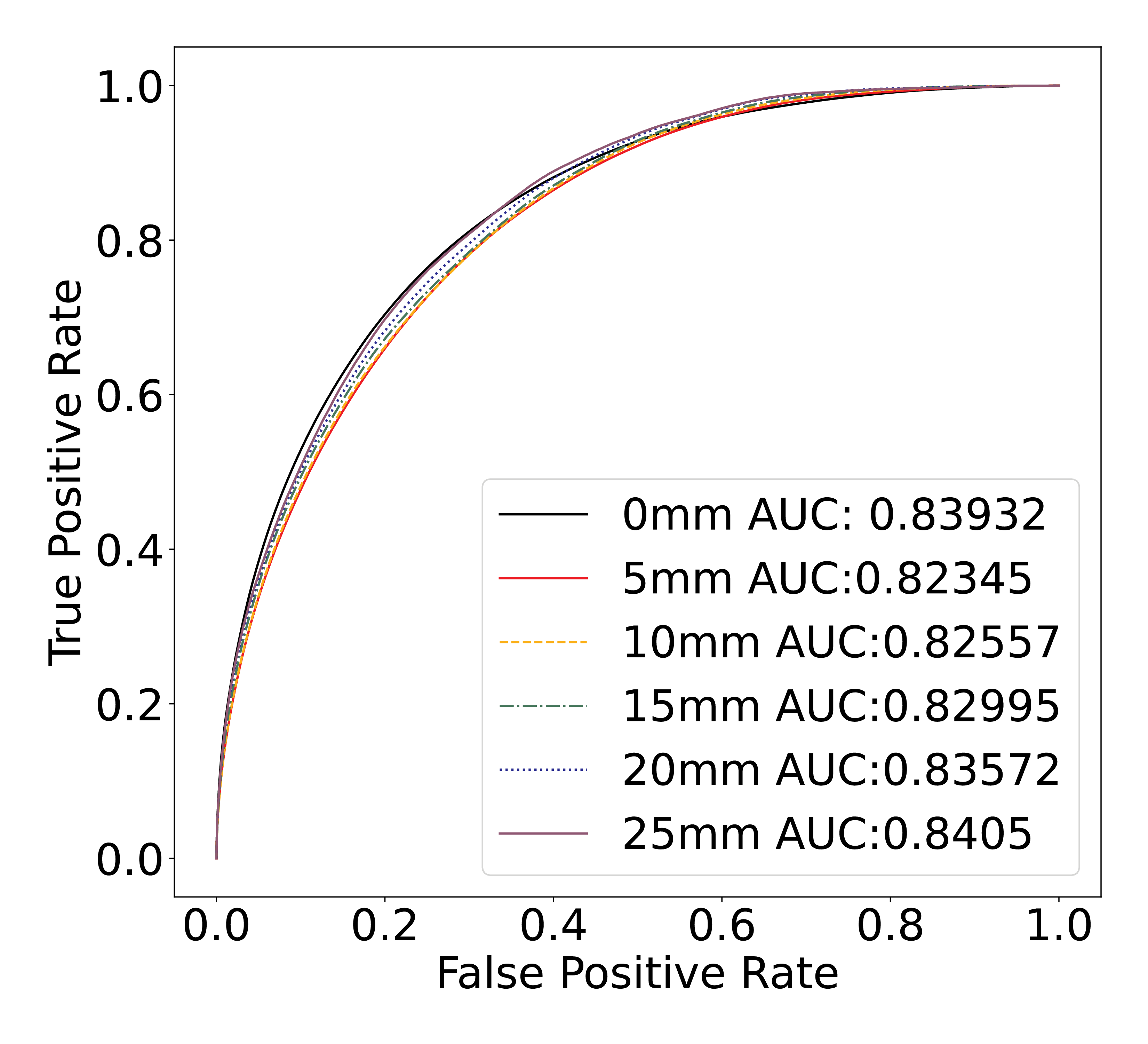}
    \caption{ConvCNP}
    \end{subfigure}
    \begin{subfigure}{0.3\textwidth}
         \includegraphics[width= \textwidth]{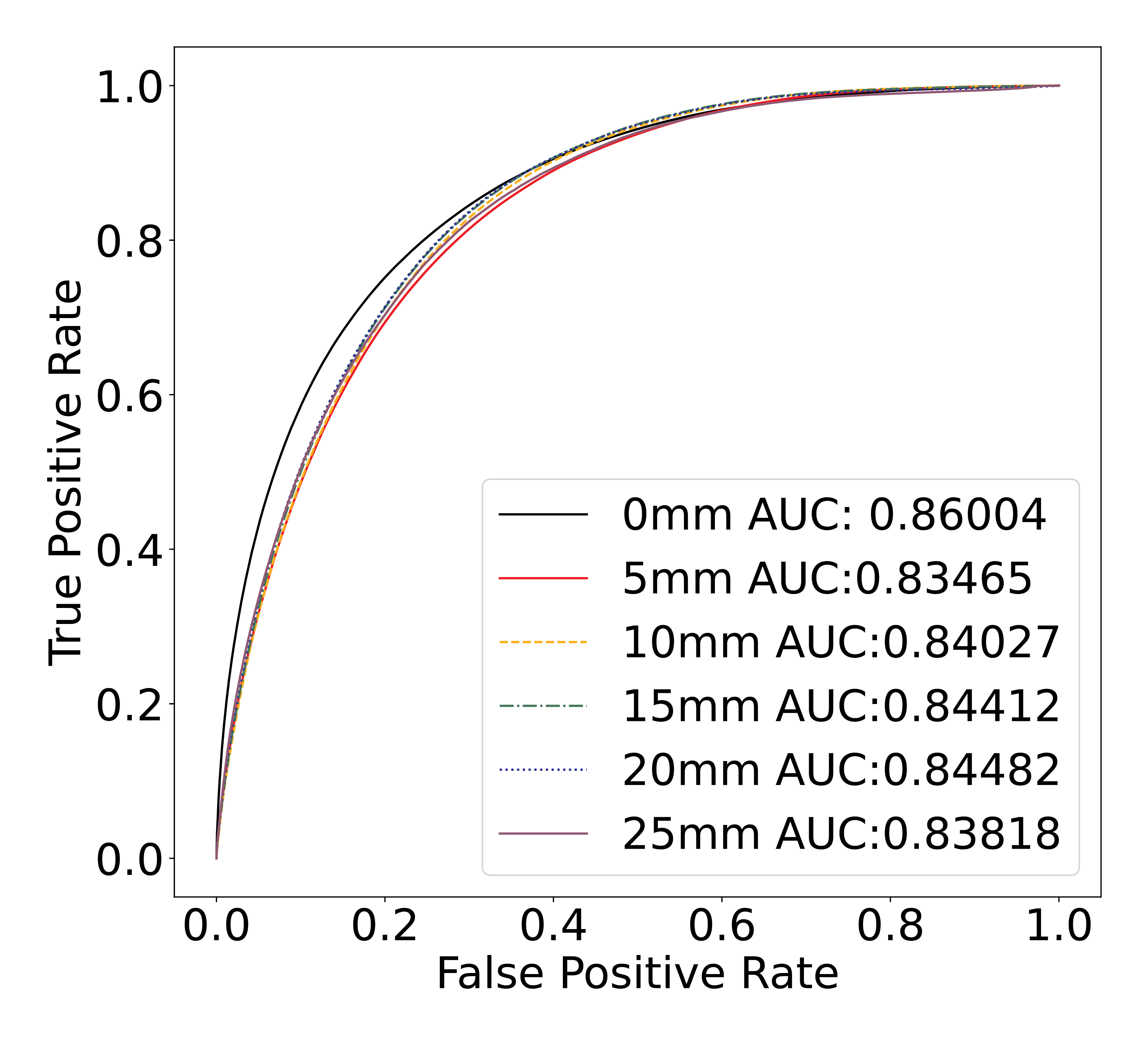}
    \caption{Cens-JGNM}
    \end{subfigure}

    \caption{\textbf{Bechmarking:} ROCs for two benchmark methods, namely the (a) the VAE-GAN and (b) ConvCNP as well as our (c) Cens-JGNM approach. Our model demonstrates the best performance, closely matched by the ConvCNP while the VAE-GAN achieves the lowest performance of the three methods.}
    \label{fig:roc_bench}
\end{figure}

\begin{figure}[ht]
\centering
        \begin{subfigure}{0.24\textwidth}
         \includegraphics[width= \textwidth]{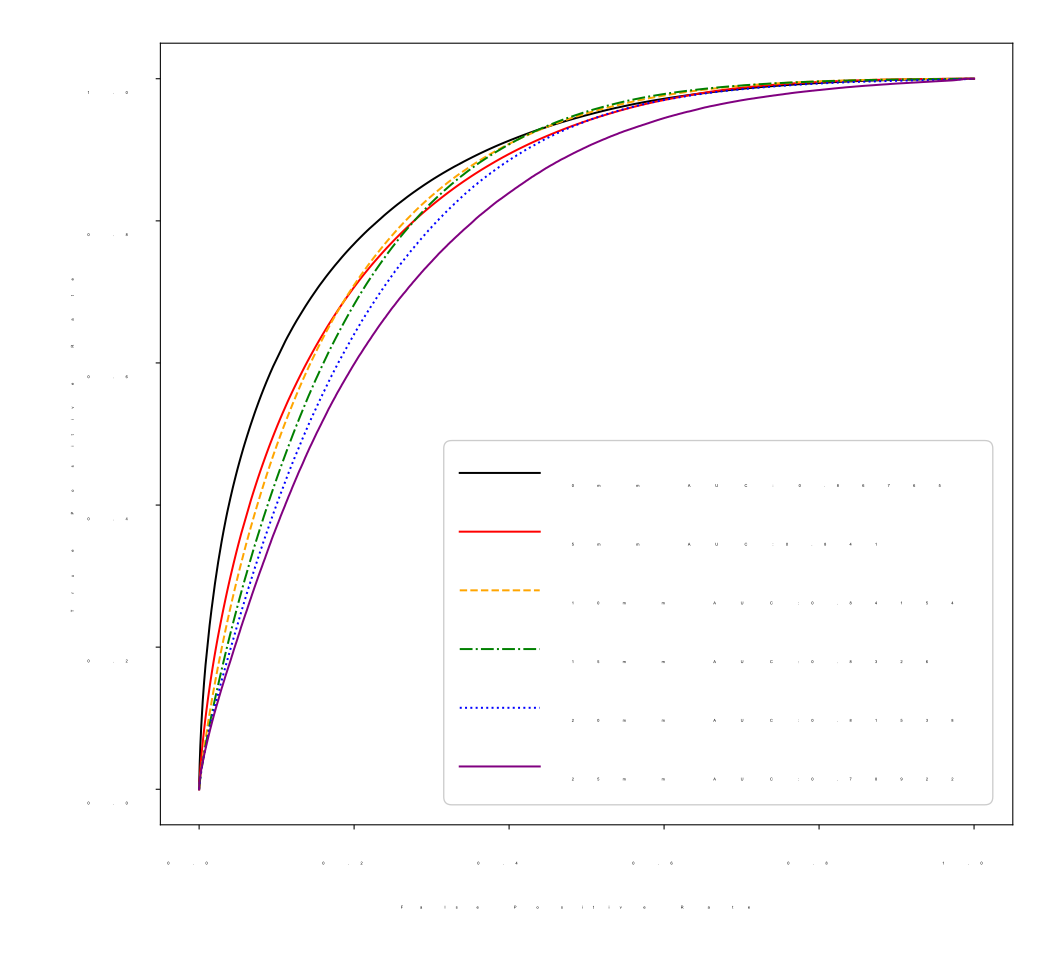}
    \caption{10 Years}
    \end{subfigure}
        \begin{subfigure}{0.24\textwidth}
         \includegraphics[width= \textwidth]{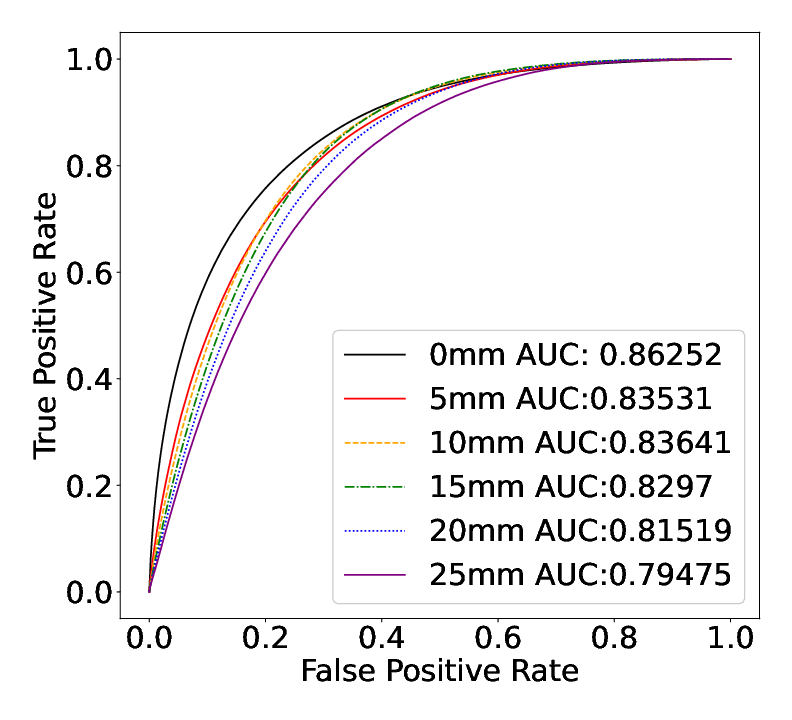}
    \caption{8 years}
    \end{subfigure}
    \begin{subfigure}{0.24\textwidth}
         \includegraphics[width= \textwidth]{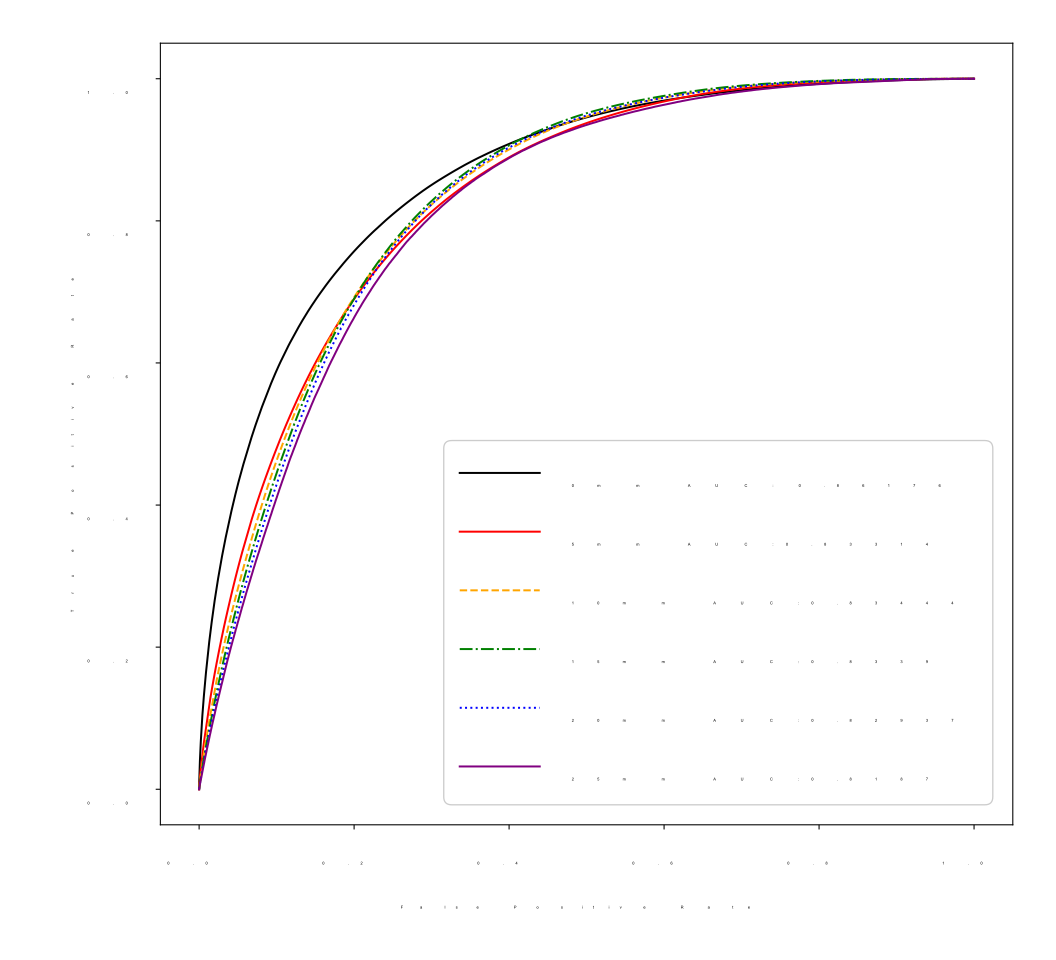}
    \caption{6 years}
    \end{subfigure}
        \begin{subfigure}{0.24\textwidth}
         \includegraphics[width= \textwidth]{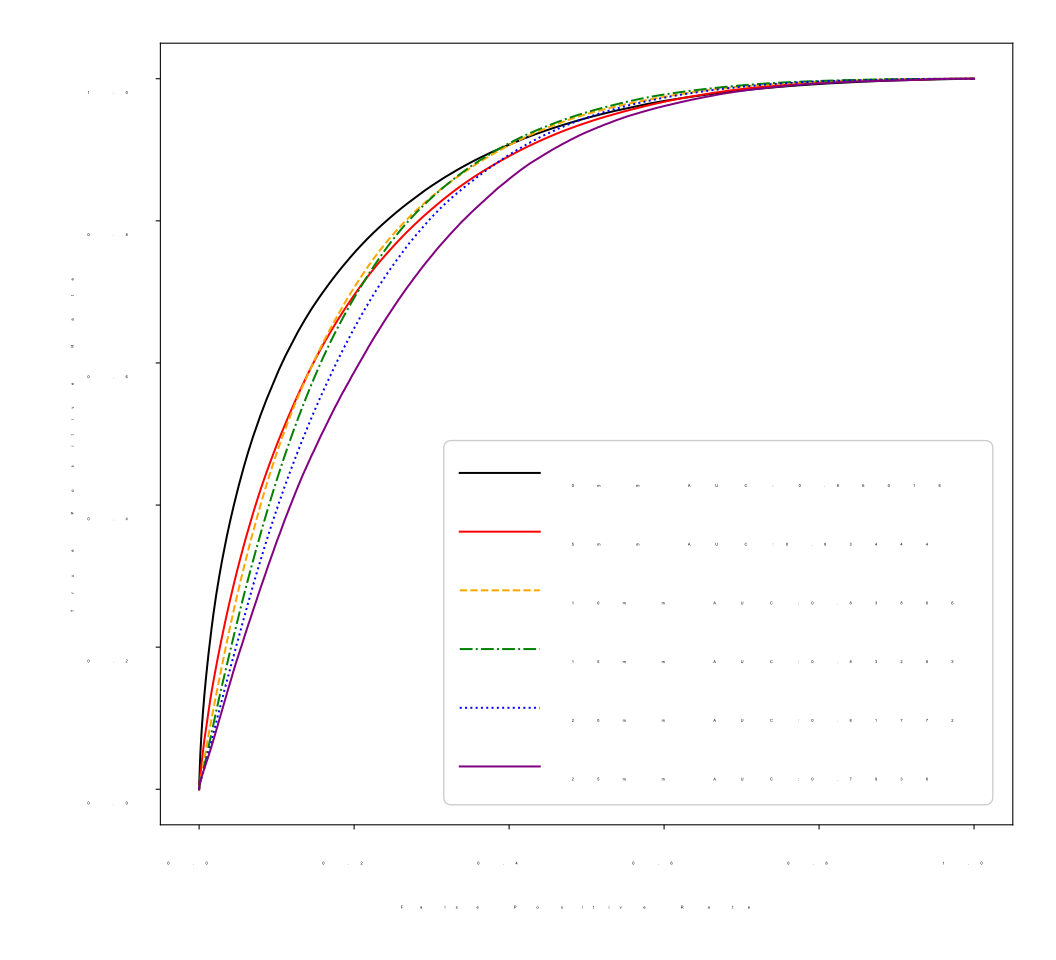}
    \caption{4 years}
    \end{subfigure}
    \caption{\textbf{Robustness:} ROCs for robustness experiments where our Cens-JGNM approach was fitted on (a) 10, (b) 8, (c) 6 and (d) 4 years of data. No clear trend can be noticed, suggesting our approach is robust to any data amount size above four years.}
    \label{fig:roc_rob}
\end{figure}

 \section{Code Reproducibility}
The code for the joint generalised neural model can be found at \url{https://github.com/Rilwan-Adewoyin/NeuralGLM}. Code for all figures shown and for the censored latent Gaussian copula can be found at \url{https://github.com/Huk-David/SpaDep_VCop/tree/4Paper}.

\end{document}